\documentclass{aa}
\usepackage{psfig,latexsym,longtable,lscape}

\def\ergsec{\hbox{erg s$^{-1}$}}
\def\ergcm{\hbox{erg cm$^{-2}$ s$^{-1}$}}

\def\et{et al. }
\def\HI{\hbox{H\,{\sc i}}}
\def\HII{\hbox{H\,{\sc ii}}}
 
\begin{document}
 
   \thesaurus{3; 
              (09.10.1; 
	       11.09.1 NGC 3079; 
               11.01.2; 
               11.19.2; 
               11.10.1; 
               13.25.2) 
             }
   \title{NGC~3079: X-ray emission from the nuclear super-bubble and halo }
 
 
   \author{W.~Pietsch\inst{1} \and
           G.~Trinchieri\inst{1,2} \and
           A.~Vogler\inst{1} 
           }
 
   \offprints{W.~Pietsch}
   \mail{wnp@mpe.mpg.de}
 
   \institute{Max Planck Institut f\"ur extraterrestrische Physik,
              Giessenbachstra\ss e, D-85740 Garching, Germany
   \and  Osservatorio Astronomico di Brera,
              via Brera 28, Milano, Italy
                     }
 
   \date{Received 6 July 1998 / Accepted 6 October 1998}
   \titlerunning{X-ray emission from NGC~3079} 
   \maketitle
 
   \begin{abstract}
We report the results of the spatial and spectral analysis of the ROSAT
HRI and PSPC observations of the edge-on spiral galaxy NGC~3079. 
We detected several sources in the field of NGC~3079 with both PSPC and
HRI, and complex emission from the inner 5\arcmin\ around NGC~3079.
We have identified possible counterparts for several of the sources outside
NGC~3079 by comparison with optical plates and catalogues. 

The X-ray emission from NGC~3079 has a L$_{\rm X} = 3\times 10^{40}$ \ergsec\
and can be resolved into the following three components:
\begin{itemize}
\item Extended emission in the innermost region, with 
L$_{\rm X} =  1\times 10^{40}$ \ergsec, 
coincident with the super-bubble seen in optical
images. The active nucleus may contribute to the emission as a point source. 
\item Emission from the disk of the galaxy, 
with L$_{\rm X} = 7\times 10^{39}$ \ergsec, that
can be partly resolved by the HRI in 3 point-like sources with luminosities of
$\sim 6 \times 10^{38}$ \ergsec\ each.
\item Very soft X-shaped emission from the halo,
with L$_{\rm X} = 6\times 10^{39}$ \ergsec, 
extending to a diameter of 27 kpc.
\end{itemize} 

The X-ray luminosity of NGC~3079 is  higher by a factor of 10 compared to 
other galaxies of similar optical luminosity and we argue that this may be 
caused 
by the presence of an AGN rather than by starburst activity. The influence
of the AGN on the companion galaxies in the NGC~3079 group are discussed.
 
   \end{abstract}

\section{Introduction}
   \begin{table}
      \caption{Parameters of NGC~3079}
         \label{parameters}
         \begin{flushleft}
         \begin{tabular}{lrr}
            \hline
            \noalign{\smallskip}
 & &Ref.  \\
            \noalign{\smallskip}
            \hline
            \noalign{\smallskip}
Type & SBm & 1 \\
            \noalign{\smallskip}
Assumed distance & 17.3 Mpc  & 2 \\
& (hence 1$'\cor5.0$~kpc) & \\
            \noalign{\smallskip}
Position of &  R.A. 10$^{\rm h}$01$^{\rm m}$57\fs75 & 3 \\
center (J2000.0) & Dec.
$\phantom{0}$55\degr40\arcmin47\farcs0 & \\
            \noalign{\smallskip}
D$_{25}$ diameter & 7\farcm7 & 1 \\
            \noalign{\smallskip}
Corrected D$_{25}$ diam. & 5\farcm5 & 1 \\
            \noalign{\smallskip}
Axial ratio & 0.22 & 1 \\
            \noalign{\smallskip}
P.A. major axis & 166\degr & 3 \\
            \noalign{\smallskip}
Inclination & 84\degr & 3 \\
            \noalign{\smallskip}
Galactic foreground N$_{\rm H}$ &0.8$\times10^{20}$~cm$^{-2}$ & 4 \\
            \noalign{\smallskip}
            \hline
            \noalign{\smallskip}
         \end{tabular}
         \end{flushleft}
{
References: \\
(1) \ \ Tully 1988 \\
(2) \ \ Tully \et 1992 \\
(3) \ \ Irwin \& Seaquist 1991  \\
(4) \ \ Dickey \& Lockman 1990 \\
}
   \end{table}
The nearby spiral galaxy NGC~3079 is seen almost edge-on and has
been investigated in detail at many wavelengths. Relevant
galaxy parameters used throughout this paper are presented in 
Table~\ref{parameters}. 
In optical light NGC~3079 shows disturbed morphology and dust lanes.
The nuclear spectra indicate LINER/Seyfert 2 activity (Heckman 1980, 
Ford et al. 1986).  
In radio, NGC~3079 stands out for anomalous filamentary and
bubble-like structure extending for $\sim$3 kpc along the minor axis to 
both sides of a bright compact nucleus 
(de Bruyn 1977, Seaquist et al. 1978, Hummel et al. 1983, 
Duric et al. 1983, Duric \& 
Seaquist 1988). Extended radio emission also arises from the disk 
of the galaxy. A 
somewhat smaller loop of optical H$\alpha$ emission (Ford et al. 
1986, Duric \& Seaquist 1988) or a super-bubble (Veilleux et al.
1994) appears associated with the eastern radio loop. 
With the help of optical long slit spectra these features 
have been interpreted as evidence of a super-wind powered by a
nuclear starburst (Heckman et al. 1990) or an active nucleus
(Filippenko \& Sargent 1992).
Veilleux et al. (1995) have further investigated the disk-halo
connection in a detailed study of the diffuse ionized medium (DIM). 
They find that DIM contributes $\sim$30\% of the total H$_{\alpha}$
emission of the galaxy disk within a radius of 10 kpc and that, within
a radius of 5 kpc, X-shaped filaments rise for more than 4 kpc 
above the disk plane. Several bubbles and filaments within 1 kpc of 
the disk plane are interpreted as direct evidence for gas flow
between the disk and halo.
Within its distance, NGC~3079 is known to be one of the brightest 
sources of FIR continuum, CO line and H$_2$O maser emission
(e.g. Henkel et al. 1984, Soifer et al. 1989, Irwin \& Sofue 1992, 
Greenhill et al. 1995, Harwarden et al. 1995, Braine et al. 1997) 
reminiscent of molecule and dust clouds and circumnuclear
starburst activity.
The \HI\ distribution has been modeled by Irwin \& Seaquist (1991), and 
slightly resolved \HI\ and OH absorption was discovered against the
radio core (Gallimore et al. 1994, Baan \& Irwin 1995) 
indicative of a central mass
within the inner 90 pc radius of more than 10$^8M_{\sun}$. 

NGC~3079 is member of a group of galaxies consisting of the giant spiral,
NGC~3079, and two small companions, MCG~9-17-9, 6\farcm5 to the NW, and 
NGC~3073 (= Mrk~131), 10$'$ to the SW (Irwin et al. 1987, Irwin \& 
Seaquist 1991). MCG~9-17-9 is a small spiral of type Sb-Sc and 
NGC~3073 an early type (SAB0-) galaxy. While in the 
optical the two companions do not show signs of disturbance, in the radio
NGC~3073 is found to exhibit an elongated \HI\ tail which is aligned with
the nucleus of NGC~3079, and the \HI\ emission of MCG~9-17-9 is slightly 
extended in the direction of the nucleus of NGC~3079, too. 
An \HII\ region spectrum coupled with deep Balmer absorption lines
indicates that the stellar population of NGC 3073 is very young.
Filippenko \& Sargent (1992 and references therein) postulate that the
copious star formation observed in NGC 3073 might have been triggered by
the super-wind from NGC 3079 (see above), which, in projection, points in
the direction of NGC 3073. From these observations it has been speculated 
by Irwin et al. that a hot inter-galaxy medium may be present in this group 
that might be observable in X-rays.

NGC~3079 was detected as an X-ray source by the Einstein observatory
(Fabbiano et al. 1982, 1992) with a luminosity of $2.1\times10^{40}$\,\ergsec\ 
in the 0.5--4 keV range (corrected to a distance of 17.8 kpc and
for Galactic foreground absorption of $8.4\times10^{19}$\,cm$^{-2}$, 
and assuming a 
thermal brems\-strah\-lung spectrum with a temperature of 5 keV).   
ROSAT PSPC data of the galaxy have been analyzed by Reichert et al. (1994)
and Read et al. (1997). They find an unresolved nuclear point source
with a luminosity in the 0.1--2 keV band of $9.5\times10^{39}$\,\ergsec\ and
a diffuse emission component of $2.1\times10^{40}$\,\ergsec. Dahlem et al.
(1998) investigate the integral galaxy spectrum using ROSAT PSPC and ASCA
data, and in addition present an overlay of ROSAT HRI contours
over an H$_{\alpha}$ image of the center of the galaxy. 

In this paper we present ROSAT HRI observation of NGC~3079 and a
detailed reanalysis of the ROSAT PSPC data extracted from the ROSAT 
archive.

\section{Observations and data analysis}
\label{observations}
NGC~3079 was observed from October 20 to 25, 1992 with the ROSAT HRI 
and from November 13 to 15, 1991
with the ROSAT PSPC, for a total observing time of 20.7 ks (HRI) and 19.0 ks
(PSPC).   The observations were  split into 11 and 9  observation intervals
(OBIs) for HRI and PSPC, respectively. To analyse the data for times of
high detector background, we investigated rates of the master veto rate 
counter for the PSPC and the invalid counts for the HRI integrated over 60 s.
The rates reach maxima of 280 cts s$^{-1}$ (82 cts s$^{-1}$), but stay 
below 200 cts s$^{-1}$ (60 cts s$^{-1}$) for 98\% (99\%) of the time for the
PSPC (HRI). We therefore decided not to reject events due to times of 
high background.

\subsection{Attitude corrections}
To improve on the attitude solution we have adopted two subsequent
techniques.  For the HRI data we analyzed individual OBIs and 
aligned the data with the help of 7 bright point 
sources (namely H1, H2, H8, H9, H15, H20, H23, cp. Sect. 3.1) to the average 
position.   While the offset in the position of these sources
was $\le$3\arcsec\ for most OBIs, for OBI 5 the offset was $\sim $12\arcsec. 
The photons were corrected for the offsets. 

We also looked for possible optical counterparts (see Appendix A.1 and
A.2) with the aid of APM
finding charts (Irwin \et 1994), and compared the optical and X-ray
positions, to determine a possible systematic error on the absolute
attitude solution.   We found optical candidates for  9 sources in the
HRI and 7 in the PSPC.    We determined a systematic
shift of 3\farcs 4 to the east and 3\farcs 9 to the north and 
an additional counterclockwise rotation of 0\fdg39 for the HRI
observation. The corrected center of 
the HRI pointing direction is $\alpha$ = $10^{\rm h}01^{\rm m}57\fs 2$,~
$\delta = 55\degr42\arcmin 39\farcs 9$ (J2000.0).  For the PSPC, we
determined a shift of 8\farcs 8 to 
the east and 6\farcs 5 to the south and an additional 
counterclockwise rotation of 0\fdg2.  The corrected center
of the PSPC pointing is $\alpha$ = $10^{\rm h}01^{\rm m}56\fs 5$,~
$\delta = 55\degr40\arcmin 41\farcs 5$ (J2000.0).
Source lists and images
have been corrected for these systematic effects. The remaining position
uncertainty is less than 3\arcsec\ for both instruments.

The attitude corrections determined above for HRI and PSPC are in good 
agreement with boresight parameters determined from a larger 
sample of observations that now are used for the SASS re-processing 
(M. K\"urster, private communication).

\subsection{Iso-intensity contour maps}
\label{maps-anal}
For the PSPC data, contour plots have been obtained  from images that
were the result of the superposition of sub-images with 5\arcsec\ bin size
in the 8 standard bands (R1 to R8, cf. Snowden et al. 1994), corrected for 
exposure, vignetting, and dead time and smoothed
with a Gaussian filter having a FWHM corresponding to the on-axis point
spread function (PSF) of that particular energy band.  The FWHM values
used range from 52\arcsec\ to 24\arcsec. 
The average background was calculated from a source free region to the
north of NGC~3079.  

For the HRI, contour plots have been obtained from images with 2\farcs5  
bin size which were corrected for dead time and smoothed with a 
12\arcsec\ FWHM Gaussian filter. 
To reduce the background due to  UV emission or cosmic rays  we used
only those events detected in the HRI raw Pulse Height Amplitude
channels 2--8.

The resulting images are discussed in Sect. 3.1. 
\subsection{Source detection}
\label{sdet-anal}
We performed source detection and position determination 
with the EXSAS local detect, map detect,
and maximum likelihood algorithms (Zimmermann et al. 1992). 
Maximum likelihood values (${\rm L}$)
can be converted into probabilities (${\rm P}$) through 
${\rm P \approx 1-e^{-L}}$,
so that ${\rm L}=8$ corresponds to a Gaussian significance of about 3.6$\sigma$
and ${\rm L}=10$ corresponds to a Gaussian significance of about
3.9$\sigma$
(Cruddace et al. 1988; Zimmermann et al. 1994).

\begin{table}
\caption{Conversion factors from count rates to fluxes (0.1--2.4 keV) 
for the ROSAT HRI and PSPC detector (broad band) in units 
of $10^{-11}$~erg~cm$^{-2}$~cts$^{-1}$, corrected for Galactic absorption}
\label{conversion}
\begin{flushleft}
\begin{tabular}{rcc}
\hline
\noalign{\smallskip}
& $\phantom{^{~\ast}}$THBR$^{~\ast}$
&$\phantom{^{~\dagger}}$THPL$^{~\dagger}$\\
Temp.& $5$~keV &$0.3$~keV \\
\noalign{\smallskip}
\hline
\noalign{\smallskip}
HRI & 3.95 & 3.56 \\
PSPC &1.16 & 1.00\\
\noalign{\smallskip}
\hline
         \end{tabular}
         \end{flushleft}
\[
\begin{array}{lp{0.95\linewidth}}
$$^\ast$$ & thermal bremsstrahlung spectrum (we assumed ${\rm T}=5$~keV
for conversion of point-like sources)\\
$$^\dagger$$ & thin thermal plasma (we assumed ${\rm T}=0.3$~keV for diffuse
X-ray emission components)
\end{array}
\]
\end{table}

\begin{figure*}
 \resizebox{12cm}{!}{\psfig{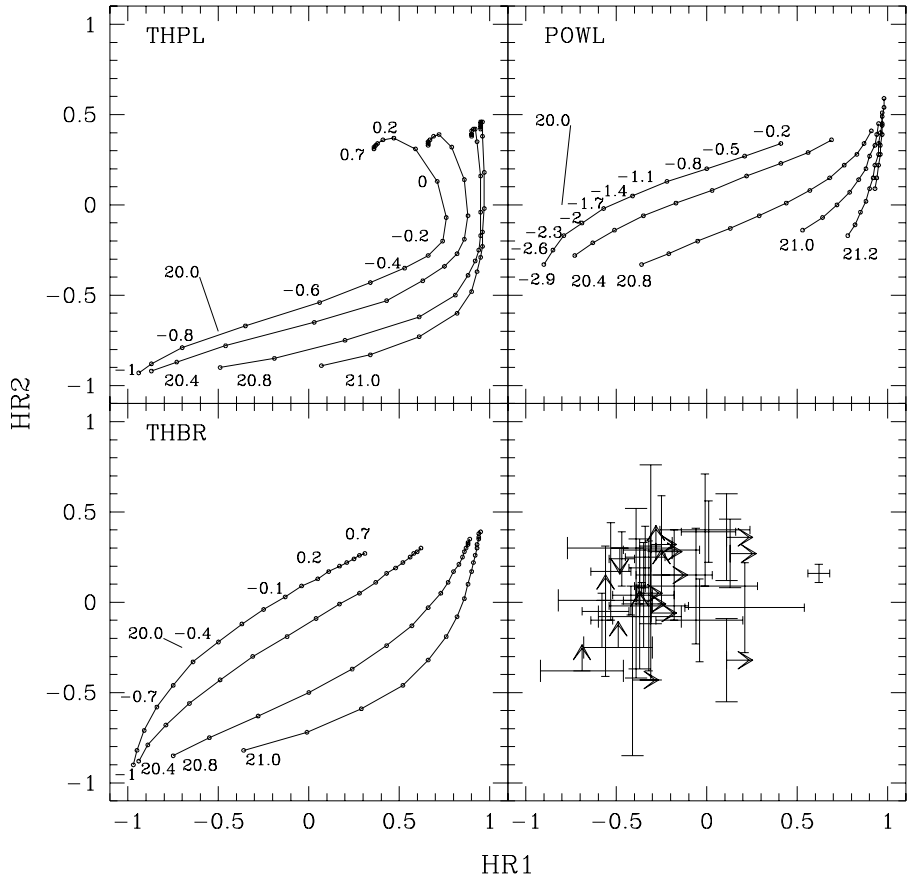}}
 \hfill
 \parbox[b]{55mm}{
   \caption{Expected distribution of HR1 and HR2 
   for different spectral models and spectral parameters and observed 
   values. Curves are drawn for different assumed equivalent
absorbing column, as given in the figure (in logarithmic value) and
for different temperatures/indexes.   
   Left panels: Curves for Raymond 
   \& Smith (top) and thermal Bremsstrahlung spectra.  Dots along the
curves indicate different temperatures, in steps of 0.1 in
log(T$_7$), where  T$_7$ is T/$10^7$ K.  Some of the points are
labeled for easy reference.  
   Upper right pannel: Curves for power law spectra.  Dots along the
curves indicate different photon index, as indicated.  
   Lower right pannel: Observed hardness ratios and limits }
   \label{hr-theor}}
\end{figure*}

Once the count rates of sources are obtained (see below for the details
on how they have been obtained in the two instruments), they are converted
into fluxes in the ROSAT band (0.1--2.4 keV) assuming a 5~keV thermal 
brems\-strah\-lung (cf. Table~\ref{conversion}). To estimate possible
errors in the X-ray luminosities due to the selection of a wrong model
or temperature, conversion factors for a 0.5~keV thermal brems\-strah\-lung
and for a 0.3~keV and a 3~keV thin thermal plasma spectrum have also
been calculated.  For this range of temperature and models, the
resulting values change by $\le$ 15\%.

Different considerations were applied to the HRI and the PSPC data to
better suit the properties of these instruments. 

\begin{description}
\item{\rm HRI}:  Sources were searched for in the inner $\sim37'$ 
diameter circle about the field's center.
Again, we used
only those events detected in the HRI raw Pulse Height Amplitude
channels 2--8 (see Sect. 2.2). Sources with a ${\rm L}\ge8$ were
accepted.

\item{\rm PSPC}: 
Sources were searched for in the inner $35\arcmin\times35\arcmin$\ field
centered on NGC~3079 in the five standard ROSAT energy bands:
``broad'' (0.1--2.4~keV),
``soft'' (0.1--0.4 keV), ``hard''
(0.5--2.0~keV),
``hard1'' (0.5--0.9~keV), and ``hard2'' (0.9--2.0~keV).
Sources for which we obtained ${\rm L}\ge10$ in at least one of the bands
were considered. We have used a slightly higher L value than in the HRI 
analysis since for the
PSPC we are already getting background limited. 

Once the existence of the source is established, its position is determined from
the band with the highest ${\rm L}$ value.  This is then used to derive
the net counts in the five standard bands defined above in an
aperture corresponding to $2.5\times$\,FWHM of the energy band
considered.  The background is always taken from the same area in the
corresponding background map (see EXSAS manual, Zimmermann et al. 1994).
These values are used to calculate
hardness ratios and their corresponding errors: 
HR1 = (hard--soft)/(hard+soft) and HR2 =
(hard2--hard1)/(hard2+hard1), where all of the quantities are corrected
for the appropriate vignetting correction (see EXSAS manual).
If the source is not detected in one of the four bands, however,
the corresponding hardness ratio cannot be simply calculated.  We
have therefore modified the algorithm, and determined to compute the
hardness ratio using the 2$\sigma$ upper limit instead of the net counts,
when the signal-to-noise ratio in a particular band is less than 2.  
This allows us to estimate a
maximum or minimum value that the hardness ratio can have.  
When both quantities are upper
limits no hardness ratio is calculated.  

Hardness ratios can be used to have a crude estimate of the spectral
parameters that best apply to the energy distribution of the source
photons, when the statistics do not allow a more detailed analysis.  To
show this we have also calculated ``theoretical" hardness ratios from the
spectral distribution of some standard spectral models ($i.e.$ 
Raymond \& Smith, power law
and thermal brems\-strah\-lung) with a low energy cut-off.  These are shown in
Fig.~\ref{hr-theor}, and can be used as a comparison to the observed
values of HR1 and HR2 (see Sects. \ref{spectra-anal} and Appendix A.1).  
\end{description}

\section{Results}
\begin{figure*}
  \resizebox{12cm}{!}{\psfig{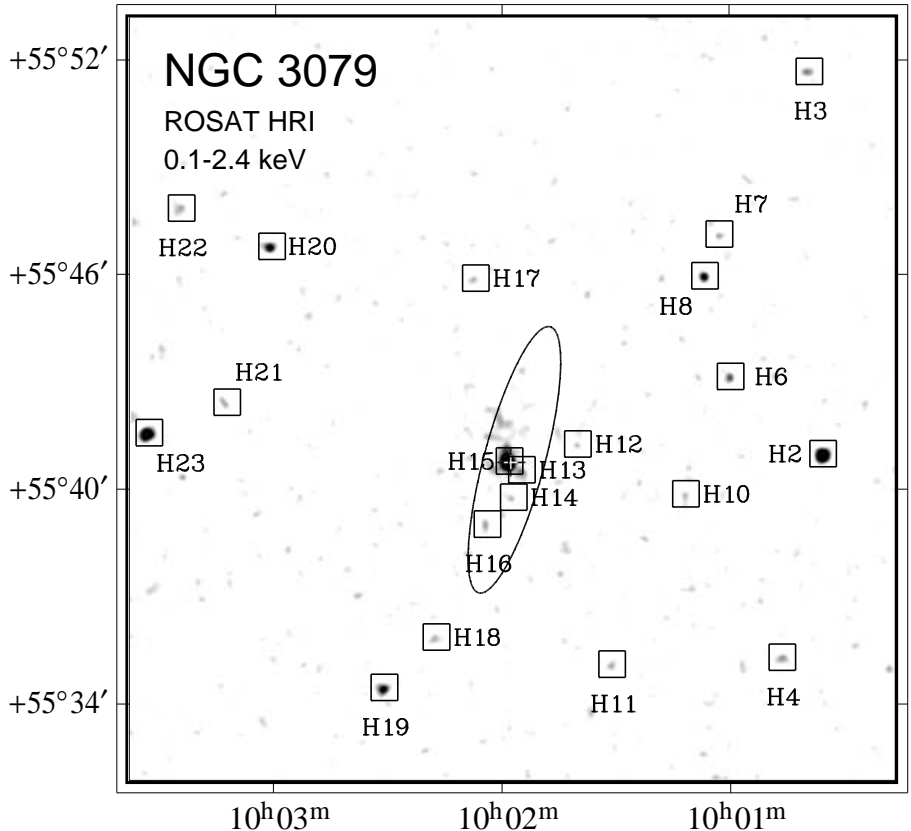}}
  \hfill
  \parbox[b]{55mm}{
   \caption[]{Grey scale plot of the inner $21\farcm3 \times 21\farcm3$\
     X-ray image seen with the ROSAT HRI. The image was constructed with
     a pixel size of 2\farcs5 and smoothed with a Gaussian of 12\arcsec\  
     (FWHM).
     The center of NGC~3079 is marked with a
     cross, the D$_{25}$ ellipse is indicated, and
     point sources (likelihood ${\rm L} \gid 8$) are enclosed by boxes 
     and numbered (see Table~\ref{sources_hri}).
     Right ascension and declination are given for J2000.0}
    \label{image_hri}}
\end{figure*}
\begin{figure*}
  \resizebox{12cm}{!}{\psfig{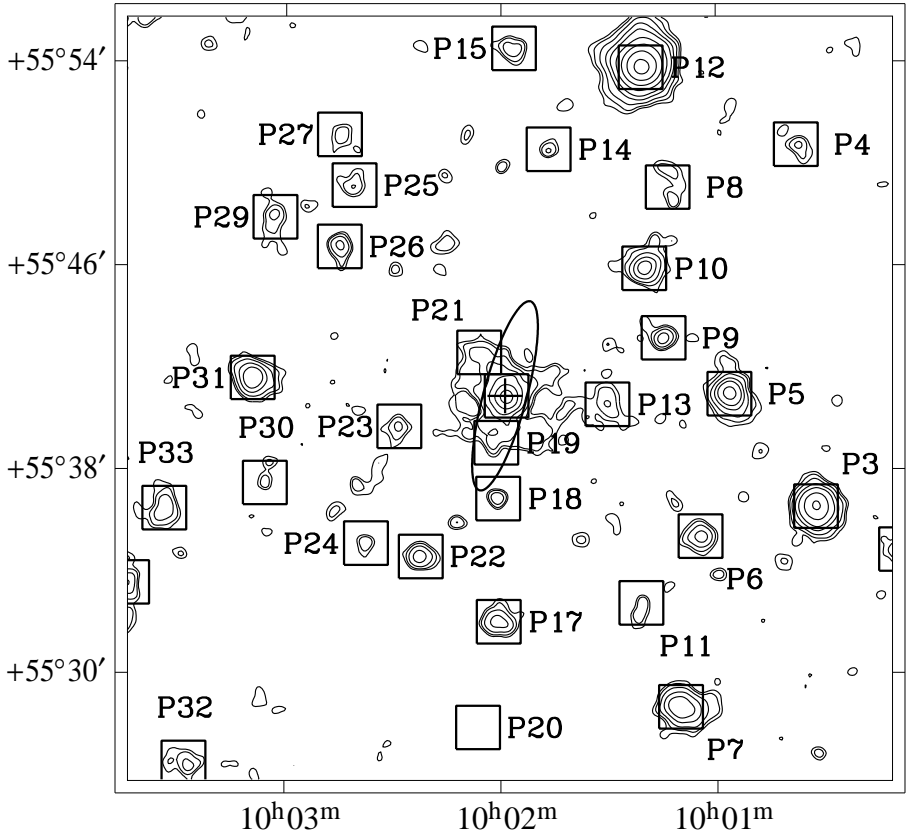}}
  \hfill
  \parbox[b]{55mm}{
    \caption[]{Contour plot of the broad band ROSAT PSPC image of the inner
     $30\arcmin\times30\arcmin$\ of the NGC~3079 field 
     (cf. Sect.~\ref{maps-anal} for how
     the image was constructed). Contours are 2, 3, 5, 9, 15, 
     31, 63, 127, and 255$\sigma$ above
     the background ($1\sigma \cor 462\times10^{-6}$\,cts~s$^{-1}$~arcmin$^{-2}$,
     background $\cor 2490\times10^{-6}$\,cts~s$^{-1}$~arcmin$^{-2}$). ROSAT PSPC
     detected sources (likelihood ${\rm L\gid 10}$) are plotted 
     as squares with source numbers written alongside (see
     Table~\ref{sources_pspc}). The position 
     of the nucleus of NGC~3079 is marked as a cross, the optical
extent is indicated by the ellipse at D$_{25}$.
     P16 detected close to the 
     galaxy center, has not been numbered}
    \label{image_pspc}}
\end{figure*}
\begin{figure*}
  \resizebox{12cm}{!}{\psfig{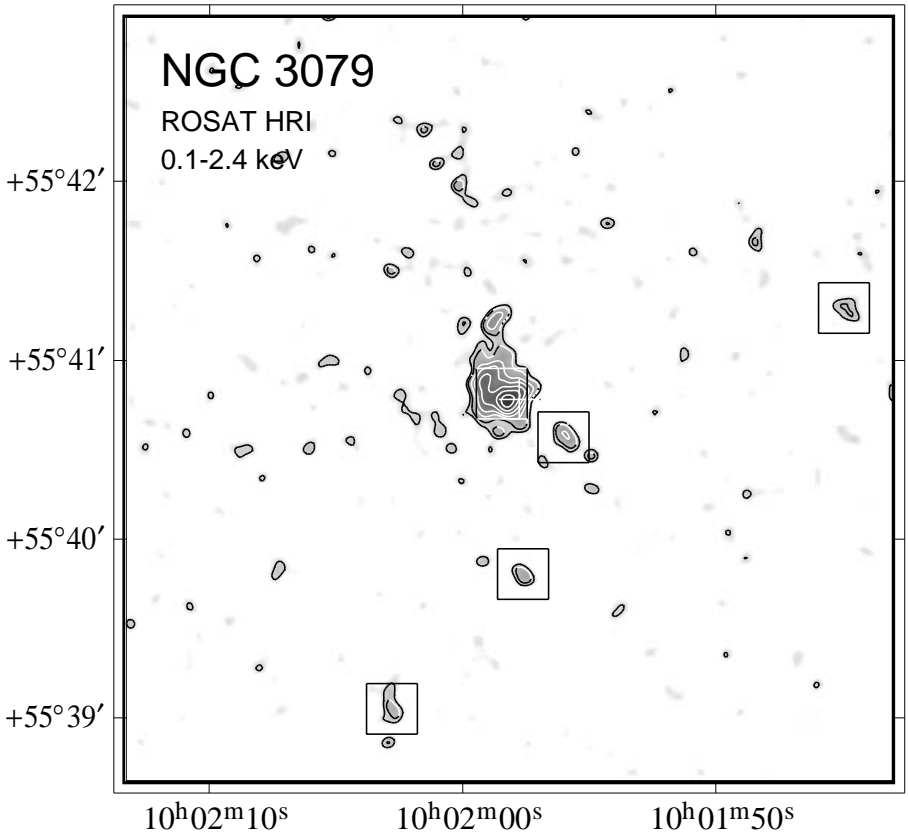}}
  \hfill
  \parbox[b]{55mm}{
    \caption[]{
     Contoured grey scale plot of the central emission region of
     NGC~3079 for ROSAT HRI.  The image is binned to a 1$''$/pixel and smoothed
     with a Gaussian function with FWHM=5$''$. 
     Contours are given in
     units of 0.5 photons accumulated per 4\farcs7 ~diameter. 
     Contour levels are 3, 5, 9, 15, 31, 63
     units.
     The center of NGC~3079 is marked by a cross,
     HRI detected point sources by squares}
    \label{master_hri_center}}
\end{figure*}
\begin{figure*}
  \resizebox{12cm}{!}{\psfig{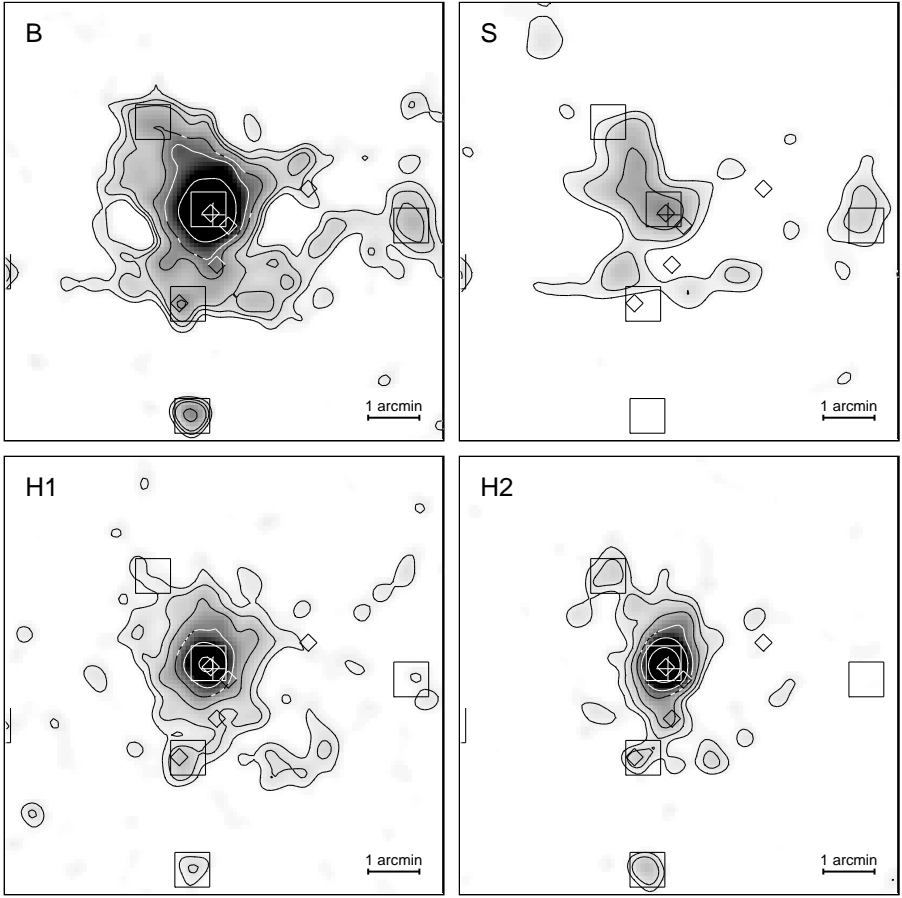}}
  \hfill
  \parbox[b]{55mm}{
    \caption[]{
     Contour plots of the central emission region of
     NGC~3079 for broad (B), soft (S), hard1 (H1), and hard2 (H2) ROSAT
     PSPC bands.
     Broad band contours are given as in Fig.~\ref{image_pspc}.
     Soft band contours are
     given in units of $\sigma$
     ($375\times 10^{-6}$\,cts~s$^{-1}$~arcmin$^{-2}$)
     above the background
     ($1905\times 10^{-6}$\,cts~s$^{-1}$~arcmin$^{-2}$),
     hard band contours (due to the negligible background
     in these bands) in
     units of 1 photon accumulated per 24\arcsec\ diameter. One
     unit$\;=418\times 10^{-6}$\,
     cts s$^{-1}$ arcmin$^{-2}$ for the hard bands.
     Contour levels are 2, 3, 5, 9, 15, 30
     units for all contour plots.
     A cross indicates the center on NGC~3079, squares indicate the positions
     of the PSPC detected sources in the field, diamonds of HRI detected sources 
     }
    \label{four_in_one}}
\end{figure*}
\subsection{Iso-intensity contour maps}
A grey scale plot of the $21\farcm3 \times 21\farcm3$\ HRI field is shown in 
Fig.~\ref{image_hri}. 
HRI sources found by the detection algorithms
(see Sec.~2.2 and 3.2)  are marked with their HRI number and 
a ellipse sketches the D$_{25}$ (see Table \ref{parameters}).

Figure~\ref{image_pspc} shows a contour map of the
$30\arcmin \times 30\arcmin$\ PSPC field centered on NGC~3079.  The positions of all
individual sources detected are marked, with their PSPC number.  Note that
P20 is detected only in the hard band, and does not appear in the
broad-band map.  The ellipse sketches the D$_{25}$.  

A smaller portion of the HRI field corresponding to the 
area of the extended emission
seen with the PSPC is presented in Fig.~\ref{master_hri_center}. The
HRI spatial resolution resolves the emission from the galaxy's plane
into a structure at the center and three individual sources, but is not
sensitive enough to show the low surface brightness emission at large
galactocentric radii. 

Figure~\ref{opt_hi_pspc} (see Sect. 4)
shows the inner part of the PSPC field superposed onto the optical image
of the galaxy.  It is
immediately apparent that the emission from NGC\,3079 is 
complex and extends both above and below the galaxy's plane. 
We have also produced maps of the emission in different energy ranges
(Fig.~\ref{four_in_one}), namely in the soft, hard1 and hard2 bands defined above.  As
can be seen by the comparisons of the iso-intensity contour maps, the softer
and harder emission show rather different morphologies.  The hard2 band
emission is aligned with the optical disk
of the galaxy, and seems to be rather confined to it, while both the
hard1
and the soft images show extensions above and below the plane.
This difference cannot be
attributed to the different response of the instrument in the different
energy bands and indicates the presence of more than one component to the
emission of NGC 3079 (see discussion later). 

\subsection{HRI/PSPC sources in the field}
\begin{table*}
\caption{X-ray properties of the
sources detected with the HRI in a $\sim$ 37\arcmin\ diameter
field centered on NGC~3079}
{
\label{sources_hri}
\begin{flushleft}
\begin{tabular}{rrrrrrrrr}
\hline
\noalign{\smallskip}
&
{\it ROSAT} name &
RA (2000) &
Dec (2000)&
R$_{err}$  &
Lik. &
Net counts &
Count rate &
PSPC\\
\noalign{\smallskip}
&
(RX~J) &
(h \ \ m \ \ s) &
(\degr \ \ \ $'$ \ \ $''$) &
($''$) &
 &
 &
(10$^{-4}$~s$^{-1}$) &
source n.\\
\noalign{\smallskip}
(1) &
(2) &
(3) &
(4) &
(5) &
(6) &
(7) &
(8) &
(9)
\\
\noalign{\smallskip}
\hline
\noalign{\smallskip}
H1 &
100032.3+553631$^*$    &
10 00 32.2&
55  36 31&
 4.4&
       201.2&
  160.2$\pm$14.2 &
     84.7$\pm$\phantom{0}7.5
&P3\\
H2 &
100056.4+554100$^{\phantom{*}}$ &
10 00 56.4&
55  41 01&
 3.5&
       167.8&
   88.1$\pm$\phantom{0}9.9 &
     44.6$\pm$\phantom{0}5.0
&P5\\
H3 &
100058.9+555141$^{\phantom{*}}$    &
10 00 59.0&
55  51 42&
 8.4&
        10.0&
   18.3$\pm$\phantom{0}5.7 &
      9.5$\pm$\phantom{0}2.9\\
H4 &
100104.7+553519$^{\phantom{*}}$    &
10 01 04.8&
55  35 19&
 8.2&
         9.9&
   17.6$\pm$\phantom{0}5.5 &
      9.1$\pm$\phantom{0}2.9
&P6\\
H5 &
100110.0+552838$^*$    &
10 01 09.6&
55  28 27&
10.9&
        18.3&
   43.9$\pm$\phantom{0}9.5 &
     23.8$\pm$\phantom{0}5.1
&P7\\[1ex]
H6 &
100114.6+554310$^{\phantom{*}}$    &
10 01 14.6&
55  43 11&
 4.1&
        23.2&
   16.8$\pm$\phantom{0}4.5 &
      8.4$\pm$\phantom{0}2.2
&P9\\
H7 &
100116.8+554709$^{\phantom{*}}$    &
10 01 16.8&
55  47 09&
 5.2&
         9.5&
   10.5$\pm$\phantom{0}3.8 &
      5.3$\pm$\phantom{0}1.9\\
H8 &
100119.7+554559$^{\phantom{*}}$    &
10 01 19.7&
55  45 59&
 3.7&
        44.6&
   25.5$\pm$\phantom{0}5.4 &
     12.7$\pm$\phantom{0}2.7&P10\\
H9 &
100120.6+555355$^{\phantom{*}}$    &
10 01 20.6&
55  53 56&
 3.1&
      2118.4&
  712.6$\pm$27.3 &
    369.9$\pm$14.1
\\
H10 &
100123.5+553953$^{\phantom{*}}$    &
10 01 23.5&
55  39 54&
 5.1&
         8.2&
    8.9$\pm$\phantom{0}3.5 &
      4.4$\pm$\phantom{0}1.7
&P12\\[1ex]
H11 &
100138.2+553508$^{\phantom{*}}$   &
10 01 38.2&
55  35 08&
 6.4&
         8.3&
   11.2$\pm$\phantom{0}4.2 &
      5.7$\pm$\phantom{0}2.1\\
H12 &
100144.9+554117$^{\phantom{*}}$    &
10 01 44.9&
55  41 18&
 4.7&
         8.4&
    8.3$\pm$\phantom{0}3.3 &
      4.1$\pm$\phantom{0}1.6\\
H13 &
100156.1+554034$^{\phantom{*}}$    &
10 01 56.2&
55  40 34&
 4.3&
        14.4&
   11.6$\pm$\phantom{0}3.8 &
      5.7$\pm$\phantom{0}1.9\\
H14 &
100157.6+553948$^{\phantom{*}}$    &
10 01 57.6&
55  39 49&
 4.4&
         9.0&
    7.7$\pm$\phantom{0}3.1 &
      3.8$\pm$\phantom{0}1.5\\
H15 &
100158.5+554049$^{\phantom{*}}$    &
10 01 58.6&
55  40 49&
 3.5&
       127.0&
  113.3$\pm$11.2 &
     55.6$\pm$\phantom{0}5.5
&P16\\[1ex]
H16 &
100202.8+553903$^{\phantom{*}}$    &
10 02 02.9&
55  39 03&
 4.5&
        10.4&
    9.1$\pm$\phantom{0}3.4 &
      4.5$\pm$\phantom{0}1.7
&P19\\
H17 &
100205.0+554555$^{\phantom{*}}$    &
10 02 05.0&
55  45 56&
 4.5&
         8.7&
    7.6$\pm$\phantom{0}3.1 &
      3.7$\pm$\phantom{0}1.5\\
H18 &
100212.9+553553$^{\phantom{*}}$    &
10 02 13.0&
55  35 54&
 5.5&
         9.4&
   11.1$\pm$\phantom{0}4.0 &
      5.6$\pm$\phantom{0}2.0\\
H19 &
100223.0+553429$^{\phantom{*}}$    &
10 02 23.0&
55  34 29&
 4.8&
        37.1&
   33.4$\pm$\phantom{0}6.6 &
     17.0$\pm$\phantom{0}3.4
&P22\\
H20 &
100245.5+554648$^{\phantom{*}}$    &
10 02 45.6&
55  46 49&
 3.9&
        49.5&
   31.4$\pm$\phantom{0}6.0 &
     15.8$\pm$\phantom{0}3.0
&P26\\[1ex]
H21 &
100254.2+554226$^{\phantom{*}}$    &
10 02 54.2&
55  42 27&
 6.9&
         8.7&
   11.5$\pm$\phantom{0}4.2 &
      5.8$\pm$\phantom{0}2.1\\
H22 &
100303.4+554752$^{\phantom{*}}$    &
10 03 03.4&
55  47 52&
 6.9&
        13.3&
   19.5$\pm$\phantom{0}5.5 &
     10.0$\pm$\phantom{0}2.8
&P29\\
H23 &
100309.6+554135$^{\phantom{*}}$    &
10 03 09.6&
55  41 35&
 4.0&
       108.8&
   72.3$\pm$\phantom{0}9.2 &
     36.9$\pm$\phantom{0}4.7
&P31\\
\noalign{\smallskip}
\hline
\end{tabular}
\end{flushleft}
\[
\begin{array}{lp{0.95\linewidth}}
$$^*$$ & The more accurate PSPC position is used to name this
source\\
\end{array}
\]
}
\end{table*}
\begin{table*}
{\footnotesize
\caption{X-ray properties of sources detected with the PSPC in a
$35\arcmin\times35\arcmin$\ field centered on NGC~3079
}
\label{sources_pspc}
\begin{flushleft}
\begin{tabular}{rrrrrrrrrr}
\hline
\noalign{\smallskip}
&
{\it ROSAT} name &
\multicolumn{2}{c}{RA \ \ \ (2000)\ \ \ Dec} &
R$_{err}$ &
Lik. &
Net counts &
Count rate &
HR1 &
HR2  \\
\noalign{\smallskip}
&
(RX~J) &
(h  \ m \ s) &
(\degr \ \ $'$ \ $''$) &
($''$) &
 &
 &
(10$^{-3}$~s$^{-1}$) &
 & \\
\noalign{\smallskip}
(1) &
(2) &
(3) &
(4) &
(5) &
(6) &
(7) &
(8) &
(9) &
(10)
 \\
\noalign{\smallskip}
\hline
\noalign{\smallskip}
 P1 &
095956.3+554639$^{\phantom{*}}$    &
09 59 56.4&
55  46 40&
17.6&
        20.4&
   74.2$\pm$13.6 &
    4.5$\pm$0.8&
-0.5$\pm$0.2&
$\ge-$0.3 \\
 P2 &
100008.8+553448$^{\phantom{*}}$    &
10 00 08.9&
55  34 49&
18.8&
        15.2&
   60.0$\pm$13.2 &
    3.6$\pm$0.8&
-0.3$\pm$0.2&
0.0$\pm$0.4\\
 P3 &
100032.3+553631$^{\phantom{*}}$    &
10 00 32.4&
55  36 31&
 4.2&
       888.7&
  640.2$\pm$28.0 &
   37.2$\pm$1.6&
-0.4$\pm$0.1&
-0.0$\pm$0.1\\
 P4 &
100037.4+555046$^{\phantom{*}}$    &
10 00 37.4&
55  50 47&
18.1&
        20.2&
   46.1$\pm$11.1 &
    2.7$\pm$0.7&
 0.2$\pm$0.3&
 0.0$\pm$0.3\\
 P5 &
100056.4+554100$^*$    &
10 00 56.4&
55  40 58&
 4.7&
       340.7&
  297.7$\pm$19.9 &
   16.7$\pm$1.1&
-0.4$\pm$0.1&
 0.2$\pm$0.1\\[1ex]
 P6 &
100104.7+553519$^*$        &
10 01 04.6&
55  35 22&
 9.5&
        40.7&
   82.0$\pm$12.7 &
    4.6$\pm$0.7&
-0.2$\pm$0.2&
 0.2$\pm$0.3\\
 P7 &
100110.0+552838$^{\phantom{*}}$    &
10 01 10.1&
55  28 39&
 6.5&
       245.4&
  288.3$\pm$20.5 &
   17.0$\pm$1.2&
-0.6$\pm$0.1&
-0.1$\pm$0.2\\
 P8 &
100113.3+554906$^{\phantom{*}}$    &
10 01 13.4&
55  49 06&
25.6&
        15.7&
   56.2$\pm$14.8 &
    3.2$\pm$0.8&
$>0.1$&
-0.3$\pm$0.2 \\
 P9 &
100114.6+554310$^{\phantom{*}}$    &
10 01 14.6&
55  43 10&
 7.0&
        62.0&
   68.8$\pm$11.3 &
    3.8$\pm$0.6&
$>0.1$&
 0.3$\pm$0.2\\
P10 &
100119.7+554559$^*$        &
10 01 19.9&
55  45 55&
 6.2&
       128.0&
  154.1$\pm$15.3 &
    8.6$\pm$0.9&
-0.3$\pm$0.1&
 0.3$\pm$0.2\\[1ex]
P11 &
100120.9+553244$^{\phantom{*}}$    &
10 01 20.9&
55  32 44&
19.5&
        12.5&
   29.2$\pm$\phantom{0}9.4 &
    1.6$\pm$0.5&
$>-0.3$&
-0.1$\pm$0.4\\
P12 &
100120.6+555355$^*$        &
10 01 20.9&
55  53 50&
 3.2&
      8066.9&
 2957.6$\pm$55.8 &
  173.1$\pm$3.3&
-0.4$\pm$0.1&
 0.1$\pm$0.1\\
P13 &
100130.2+554033$^{\phantom{*}}$    &
10 01 30.2&
55  40 34&
25.7&
        11.5&
   57.4$\pm$13.8 &
    3.1$\pm$0.8&
-0.7$\pm$0.2&
$>-0.4$\\
P14 &
100146.6+555036$^{\phantom{*}}$    &
10 01 46.6&
55  50 36&
14.7&
        10.2&
   33.1$\pm$\phantom{0}9.3 &
    1.9$\pm$0.5&
$>-0.4$&
0.1$\pm$0.5\\
P15 &
100156.3+555434$^{\phantom{*}}$    &
10 01 56.4&
55  54 35&
12.4&
        24.1&
   61.5$\pm$11.7 &
    3.6$\pm$0.7&
-0.4$\pm$0.2&
$ >-0.1$\\[1ex]
P16 &
100158.5+554049$^*$        &
10 01 58.3&
55  40 53&
 4.7&
      1022.0&
  643.4$\pm$28.5 &
   35.2$\pm$1.6&
 0.6$\pm$0.1&
 0.2$\pm$0.1\\
P17 &
100200.5+553200$^{\phantom{*}}$    &
10 02 00.5&
55  32 00&
 9.0&
        67.0&
  113.6$\pm$14.0 &
    6.4$\pm$0.8&
-0.5$\pm$0.1&
 0.2$\pm$0.3\\
P18 &
100200.7+553651$^{\phantom{*}}$    &
10 02 00.7&
55  36 51&
 9.6&
        20.5&
   44.0$\pm$\phantom{0}9.9 &
    2.4$\pm$0.5&
$>-0.3  $&
$>0.1$\\
P19 &
100202.8+553903$^*$        &
10 02 01.2&
55  39 03&
11.9&
        25.3&
   37.2$\pm$\phantom{0}9.9 &
    2.0$\pm$0.5&
0.0$\pm$0.2&
-0.1$\pm$0.2\\
P20 &
100206.2+552750$^{\phantom{*}}$    &
10 02 06.2&
55  27 50&
21.8&
        10.4&
   16.4$\pm$\phantom{0}8.7 &
    1.0$\pm$0.5&
$>-0.4     $&
0.0$\pm$0.4\\[1ex]
P21 &
100206.0+554236$^{\phantom{*}}$    &
10 02 06.0&
55  42 36&
14.9&
        14.6&
   46.0$\pm$10.8 &
    2.5$\pm$0.6&
-0.2$\pm$0.2&
0.3$\pm$0.3\\
P22 &
100223.0+553429$^*$        &
10 02 22.3&
55  34 34&
 8.2&
        51.5&
   88.4$\pm$12.5 &
    4.9$\pm$0.7&
-0.3$\pm$0.2&
0.0$\pm$0.3\\
P23 &
100228.2+553941$^{\phantom{*}}$    &
10 02 28.3&
55  39 41&
15.3&
        10.7&
   37.7$\pm$10.1 &
    2.1$\pm$0.6&
$<-0.2  $&
$---$\\
P24 &
100237.4+553505$^{\phantom{*}}$    &
10 02 37.4&
55  35 05&
14.8&
        10.4&
   30.9$\pm$\phantom{0}9.4 &
    1.7$\pm$0.5&
$>-0.4   $&
-0.4$\pm$0.4\\
P25 &
100240.8+554910$^{\phantom{*}}$    &
10 02 40.8&
55  49 11&
16.4&
        12.0&
   42.0$\pm$10.7 &
    2.4$\pm$0.6&
-0.5$\pm$0.3&
$<0.3$\\[1ex]
P26 &
100245.5+554648$^*$        &
10 02 44.9&
55  46 46&
 8.4&
        44.4&
   65.9$\pm$11.3 &
    3.7$\pm$0.6&
$ >0.1    $&
 0.4$\pm$0.2\\
P27 &
100244.9+555110$^{\phantom{*}}$    &
10 02 44.9&
55  51 11&
17.9&
        10.0&
   33.5$\pm$10.1 &
    1.9$\pm$0.6&
$ >-0.3   $&
$ >0.3    $\\
P28 &
100245.1+555758$^{\phantom{*}}$    &
10 02 45.1&
55  57 59&
12.9&
       111.3&
  241.9$\pm$21.3 &
   14.9$\pm$1.3&
0.0$\pm$0.2&
 0.4$\pm$0.2\\
P29 &
100303.4+554752$^*$        &
10 03 02.9&
55  47 55&
20.1&
        15.1&
   45.5$\pm$11.1 &
    2.6$\pm$0.6&
-0.1$\pm$0.3&
0.1$\pm$0.3\\
P30 &
100305.5+553727$^{\phantom{*}}$    &
10 03 05.5&
55  37 28&
13.3&
        16.4&
   24.2$\pm$\phantom{0}8.7 &
    1.4$\pm$0.5&
$>-0.3$&
0.3$\pm$0.4\\[1ex]
P31 &
100309.6+554135$^*$        &
10 03 09.1&
55  41 36&
 5.8&
       207.8&
  221.9$\pm$17.8 &
   12.6$\pm$1.0&
-0.5$\pm$0.1&
$>0.0$\\
P32 &
100327.8+552625$^{\phantom{*}}$    &
10 03 27.8&
55  26 26&
28.7&
        10.9&
   51.7$\pm$13.0 &
    3.2$\pm$0.8&
-0.6$\pm$0.3&
$>0.0$\\
P33 &
100333.5+553627$^{\phantom{*}}$    &
10 03 33.6&
55  36 27&
14.0&
        29.8&
   79.1$\pm$13.3 &
    4.7$\pm$0.8&
-0.6$\pm$0.1&
-0.1$\pm$0.4\\
P34 &
100343.6+553331$^{\phantom{*}}$    &
10 03 43.7&
55  33 31&
16.9&
        24.8&
   68.4$\pm$12.5 &
    4.2$\pm$0.8&
0.0$\pm$0.3&
0.4$\pm$0.3\\
\noalign{\smallskip}
\hline
\end{tabular}
\end{flushleft}
\[
\begin{array}{lp{0.95\linewidth}}
$$^*$$ & The more accurate HRI position is used to name this source \\
\end{array}
\]
}
\end{table*}
The source detection procedure yielded 23 sources in the HRI and 34 in
the PSPC above the selected likelihood threshold
for source existence.  These numbers reduce to 20 sources in the inner 
$21\farcm3\times21\farcm3$ HRI field
(Fig.~\ref{image_hri}) and 30 sources
in the inner $30\arcmin\times30\arcmin$\ of the
PSPC image (Fig.~\ref{image_pspc}).

The X-ray properties
of the sources are summarized in Tables~\ref{sources_hri} and
\ref{sources_pspc}:
source number (col. 1), ROSAT name (col. 2), right
ascension and declination (col. 3, 4), error of the source position
(col. 5, including the  3\arcsec\ systematic error for the attitude
solution), likelihood of existence (col. 6), net counts and error for
the 0.1--2.4~keV ROSAT band (col. 7), count rates and error after
applying dead time and vignetting corrections (col. 8).  For PSPC
sources we also list hardness ratios HR1 and HR2 with their relative
errors (col. 9, 10).  For PSPC sources,  positions and maximum
likelihood values have been determined from the energy band with the
highest detection likelihood, but  the count rates refer to the broad
band.  To distinguish HRI from PSPC sources, a H or a P has been
prefixed to the number for HRI or PSPC, respectively.  For sources
detected in both instruments, the ROSAT names have been derived from
the detection with the smaller error radius. These were mainly HRI
detections. Only for two sources (H1 and H5) the HRI position errors
are bigger than the corresponding PSPC errors due to source variability
or big off-axis angle (see Sect. A.1).

Only one HRI source (H15, at the center of NGC~3079) and two
PSPC sources (P16, at the center of NGC~3079, and P28) were
flagged as extended by the maximum likelihood detection algorithm.  

Besides for the galaxy's center, only one other 
PSPC source, P19, and 3 HRI sources, H13, H14, and H16, are
positioned within the D$_{25}$ contours of the galaxy.
However it is likely that also sources H12, H18, P18, and P21, are related to
NGC~3079, and they will be regarded as such in what follows (see Sect.
\ref{in3079}).  
Moreover, it is also possible that some of these sources, located in
this complex area where extended emission is also seen, are spurious
detections picked up by the detection algorithms as a consequence of a
bad background model due to the more diffuse component and represent
local enhancements.  One HRI source (H6) and the corresponding PSPC source P9 
are identified with the companion galaxy MCG~9-17-9 (see
Sect. \ref{companion_results}). Properties of other 
sources outside the D$_{25}$ ellipse of NGC~3079 are discussed 
in the appendix.

\subsection{The emission from NGC~3079}
\label{in3079}
Complex emission partially filling the D$_{25}$ ellipse 
and extending into the halo along the minor axis is detected
(see Figs.~\ref{image_hri}, \ref{image_pspc}, 
\ref{master_hri_center}, and \ref{four_in_one}).
On top of this emission, the central nuclear region and five sources 
(H12, H13, H14, H16/P19, P21) are resolved.  Two additional sources 
(P18, and H18) positioned outside the D$_{25}$ diameter are within the
\HI\ envelope of NGC~3079, and  are also
probably associated to the emission of NGC 3079. 
The properties of the 
nuclear source (H15/P16) are further investigated in 
Sect.~\ref{results_nucleus}.
The two sources H13 and H14 detected with the HRI are too close to the bright 
nucleus (25\arcsec\ 
and 60\arcsec, respectively) to be resolved by the PSPC. 
The HRI source detection algorithm did not separate a source at the 
northern end of the diffuse central emission (distance 25\arcsec),
even though the contour map of Fig.~\ref{master_hri_center} suggests
a separate peak. 
Sources P13 and H10 appear to be within the outermost PSPC
contour of Fig.~\ref{image_pspc}.  While at the present time it is not
possible to exclude the possibility that these  are unrelated 
background sources, the evidence of excess emission in this region
suggests that maybe these are the peaks of a more extended emission
probably connected with the galaxy or with the group.  
We therefore will discuss these two sources as both truly individual
sources and as a more diffuse component.  

\begin{table}
\caption{X-ray parameters of NGC~3079 sources 
}
\label{n3079_sources}
{
\begin{flushleft}
\begin{tabular}{rrrrrrr}
\hline
\noalign{\smallskip}
\multicolumn{2}{c}{Number} & $\Delta$ &Flux &Flux 
&L$_{\rm X}$ &L$_{\rm X}$ \\
\multicolumn{2}{c}{HRI PSPC} &
 &
HRI &
PSPC & 
HRI &
PSPC \\
\noalign{\smallskip}
& &($''$)
&$^\ast$ &$^\ast$ &$\ast\ast$ & $\ast\ast$ \\
\noalign{\smallskip}
(1) & (2) & (3) & (4) & (5) & (6) & (7) \\
\noalign{\smallskip}
\hline
\noalign{\smallskip}
H12 &     & & 1.6$\pm$0.6 & $<3.4$ & $5.8$ & $<15$ \\
H13 &     & & 2.2$\pm$0.7 & & $8.1$ & \\
H14 &     & & 1.5$\pm$0.6 & & $5.4$ & \\
H15 & P16 & 5.0 & 22.0$\pm$2.2 & 40.8$\pm$1.9 & $78.8$ & $146.2$ \\
    & P18 & & $<2.6$ & 2.8$\pm$0.6 & $<9.2$ & $10.0$ \\
H16 & P19 & 15.0 & 1.8$\pm$0.7 & 2.3$\pm$0.6 & $6.4$ & $8.3$ \\
H18    & & & 2.2$\pm$0.8 & 2.0$\pm$1.0 &8.1&7.1\\
    & P21 & & $<1.8$ & 2.9$\pm$0.7 & $<6$ & 10.4 \\
\noalign{\smallskip}
\hline
\end{tabular}
\end{flushleft}
\[
\begin{array}{lp{0.95\linewidth}}
$$^\ast$$ & fluxes in units of 10$^{-14}$~erg~cm$^{-2}$~s$^{-1}$
for a 5~keV thermal brems\-strah\-lung spectrum in the 0.1--2.4~keV
band, corrected for Galactic absorption\\
$$^{\ast\ast}$$ &luminosity in units of 10$^{38}$~erg~s$^{-1}$ 
assuming a distance of 17.3 Mpc.  Fluxes and luminosities (for
which errors are not quoted) of these sources, in particular for the
PSPC data, should be taken with caution, since
extended diffuse emission is included in the flux of the individual
sources due to  the relatively large dimension of the instrument point 
spread function (see text) 
\end{array}
\]
}
\end{table}

Table~\ref{n3079_sources} 
summarizes PSPC and HRI count rates, X-ray fluxes f$_{\rm X}$ and luminosities 
L$_{\rm X}$ of the sources in NGC~3079.  
For two sources that were detected only in the PSPC, and for 2 detected
in the HRI only, a 2$\sigma$ limit to the HRI (PSPC) count rates at the
same positions are calculated.  These are estimated from circles of
radii of $0.75\times$\ FWHM of the source at the off-axis distance (PSPC and HRI,
respectively).  The fluxes are then corrected by a factor 2 to
compensate for the small aperture used to estimate the net counts.  
No equivalent limit is given for H13 and H14, 
since they would not be resolved in the PSPC.

While the luminosity of the nuclear source is above $8\times
10^{39}$\,\ergsec\ most of the other sources are close to the detection limit 
and show luminosities in the range $(5 - 12) \times 10^{38}$\,\ergsec. 

The count rates of the individual sources in NGC~3079 are too low to be
used to study time variability within each individual PSPC or HRI 
observation (day time scale). We can however investigate 
time variability of the sources on one year time scale by comparing the 
PSPC and HRI observations that were taken $\sim$ 1 year apart.  
No source variability can be claimed for the sources in NGC 3079. 

The $2\sigma$ HRI upper limit for source P21 however
appears to be significantly lower than the PSPC flux.  
P21 is detected from the
algorithm in the hard2 band only, as can also be seen by the maps in
Fig.~\ref{four_in_one}.  While it is at the moment unclear whether this
should be considered a real source, or rather a local enhancement in the
diffuse emission that extends to the NE of the galactic plane, it is
clear that a variability study is severely hampered by the presence of this
latter component, given the widely different spatial resolutions and
sensitivity to low surface brightness components of the
HRI and the PSPC.  Therefore the much lower HRI flux could be in part
(totally) due to the different amounts of diffuse component in the
detection cell.   In fact, when we estimate the background locally,
namely from the 
average surface brightness of the emission at the same radial distance
from the center, the net count rate above the extended emission reduces
to almost a half, and the flux 
f$_{\rm x} \sim (1.7 \pm 0.7) \times 10^{-14}$\,\ergcm, 
comparable to the upper limit determined from the HRI data.  

\subsubsection{Radial distribution of the emission}
\begin{figure}
\psfig{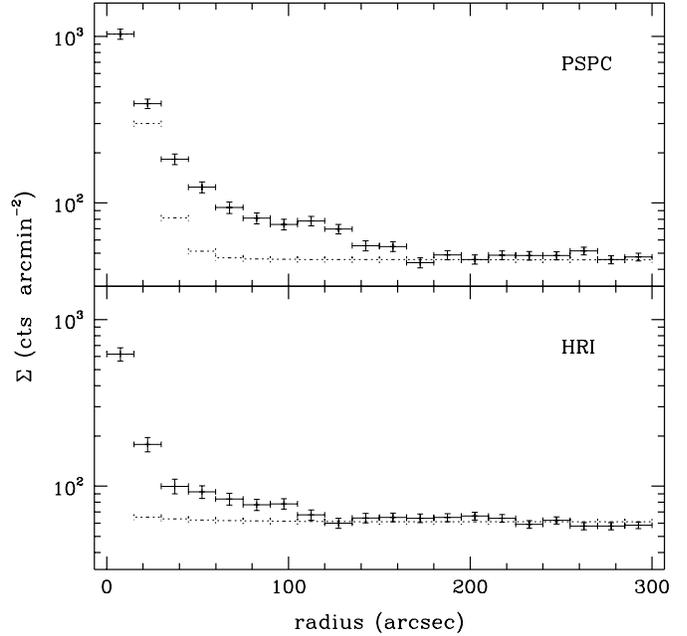}
\caption[]{Radial distribution of the detected photons,
azimuthally averaged in concentric annuli of 15$''$ width. The dashed
profiles indicates the radial distribution of a point source and the
background level estimated as explained in the text. Point sources
detected have not been removed from these profiles }
\label{radplot} 
\end{figure}
To determine the extent of the emission and the spatial distribution of
the detected photons we have produced radial surface brightness plots
from the PSPC and from the HRI data, centered at the X-ray central peak
(sources P16 or H15).  The plots are shown in Fig.~\ref{radplot}.  
The radial distribution of a point source, at the central position, binned
as the data and normalized to them in the innermost bin, is also plotted
for comparison.  In the PSPC data, the point source is simulated 
separately in energy bins of 0.1 keV which are then normalized to the
count rates in the relevant sub-band and co-added.
In the HRI data, it is obtained from the
analytical formula given in David et al. (1994) for a source on-axis.  
In both HRI and PSPC data,  the overall
photon distribution is inconsistent with a point source (however, see
later for further analysis of the HRI data).  

The azimuthally averaged surface brightness distribution of the emission
(Fig.~\ref{radplot})  
extends out to a radius of r $\sim2\farcm7$ (13.5 kpc) 
in the PSPC data, outside of
which the profile becomes constant with radius and consistent with the
background map created from the data (see Sect.~\ref{sdet-anal}).   
Similar plots in the soft and hard band indicate maximum
radii of comparable values.  In the HRI, the profile flattens at a
radius r $\sim 2'$ (10 kpc).  For the HRI,  
we can therefore determine the field background
from a region outside of the galaxy's emission by choosing an annular
region around the galaxy of 4$'-6'$
inner and outer radii, respectively.   
Point sources that lie in the background regions have not been included
for background estimates by masking them out with circles of
30\arcsec\ and 15\arcsec\ radii for PSPC and HRI, respectively.   
Correction for vignetting is negligible at these off-axis angles. 

\begin{figure}
\psfig{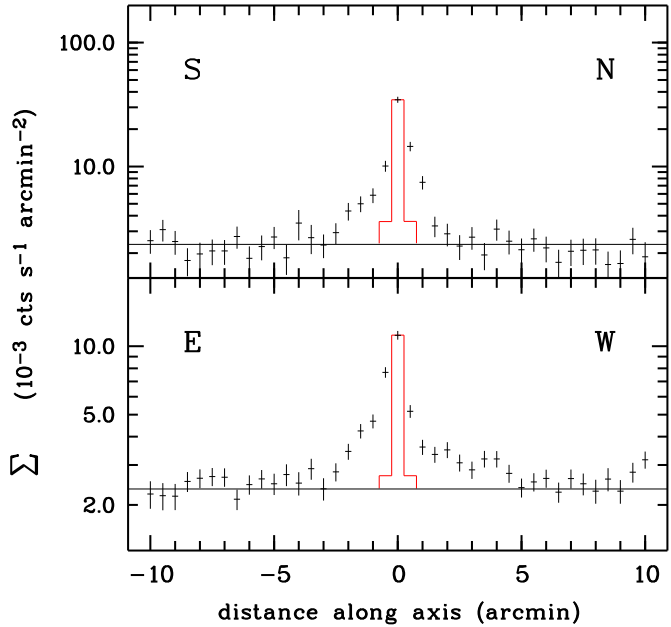}
\caption[]{
Spatial distributions of the surface brightness along the
major (above) and minor (below) axis of NGC~3079. ROSAT PSPC counts are
integrated in boxes of $30\arcsec\times60\arcsec$\ along the major axis,
covering the galaxy disk region, and in boxes of $30\arcsec\times300\arcsec$\
along the minor axis, covering the galaxy halo region. They are
centered at the distance given on the
X axis relative to galaxy's nucleus.
All PSPC sources (except P13, 
P16, P19, and P21 probably connected to the galaxy) have been cut out 
with a cut radius of
$2\times$\,FWHM of PSF at 0.3 keV. Response of a point source at the nucleus
of NGC~3079 is given as dotted histogram (normalized to 
the count rate of the central box)
}
\label{rad_maj_min}
\end{figure}
\begin{figure}
\psfig{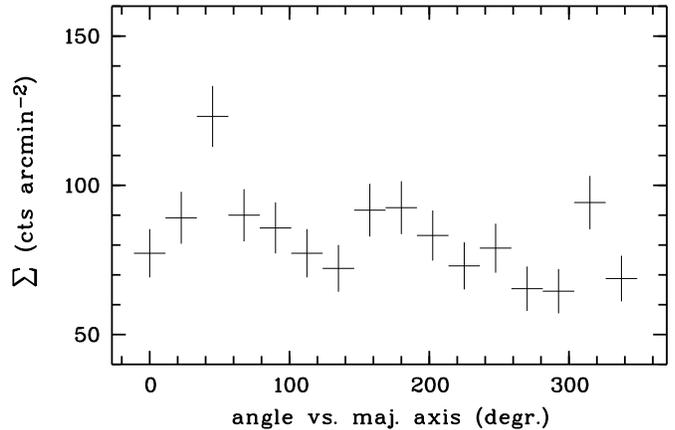}
\caption[]{
Azimuthal distributions of the surface brightness of NGC~3079. 
ROSAT PSPC counts are integrated 
over sectors of 22.5 degree within a radius of 2\farcm5. 
The central source has been cut out with 30 \arcsec\ 
radius. The sector at 0\degr\ is centered at the direction of the major axis 
and angle is counted from north to the east}
\label{azim}
\end{figure}

Given the presence of P13 in the PSPC data, and of the apparent
connection between this source and the galaxy (see
Fig.~\ref{opt_hi_pspc}), we have further analyzed the PSPC data by looking
at the radial distribution of photons in different directions and in
comparison to the expected pure field background, whose shape is
represented by the exposure map.  In fact, while the background map is
constructed from the data and therefore takes partially into account
any diffuse emission present in the field, the exposure map should
represent the PSPC response to a flat, constant radiation, while taking
into account both exposure and vignetting.  When properly normalized to
the data, the expected field background can therefore be estimated from
it (see also Trinchieri et al. 1994, Kim \& Fabbiano 1995).  
Figure~\ref{rad_maj_min} shows the results
of the comparison between the spatial profiles in different directions
relative to the exposure map.  These have been obtained by calculating
surface brightness profiles along the major and minor axis using boxes of
$30\arcsec\times60\arcsec$\ and $30\arcsec\times300\arcsec$, respectively, 
perpendicular to the axis. Sources P5, P6,
P14, P17, P18, P22, P23 and P31 
were cut out with a cut radius of $2\times$\,FWHM of PSF at 0.3 keV.
We normalized according to box area and exposure and corrected for
vignetting and dead time. The normalization of the exposure map is 
determined at 8\arcmin\ to 10\arcmin\ offset from the galaxy.

The profiles along the major and minor axes are clearly more extended
than a point source (cf. Fig.~\ref{rad_maj_min}). 
The extent along the major axis ($\sim$2\farcm5\ corresponding to 12.5 kpc 
to  both sides of the nucleus) is similar to the optical 
(about the corrected D$_{25}$). If the X-ray emission only originated from the
galaxy disk, given the galaxy's inclination one would expect a 
$\le$1\arcmin\ extent 
to both sides of the nucleus along the minor axis. Figure\ \ref{rad_maj_min}
instead shows an extension comparable to that 
along the major axis and possibly more.   In the Western direction, 
excess emission is detected out to $\sim 6'$ (30 kpc),  while only a marginal
excess is seen at distances greater than 2\farcm5 (12.5 kpc) in the Eastern
direction.  

We can quantify this excess by noticing that $\sim 221\pm 32$ net
excess counts are found in the region from 2\farcm5 -- 6\arcmin\ in the
western direction, significantly higher than the expected contribution
from the single source P13 ($\sim 57$ counts, from
Table~\ref{sources_pspc}) and also much more extended than expected
from a single point-like source.  While with the present data we cannot
exclude that P13 is indeed an individual source, the excess found
points towards interpreting it as a local enhancement onto a somewhat
irregular emission.  No optical counterpart can be seen in the 
finding charts (see Appendix A.2). The excess in the east is $\sim
49\pm 26$ net excess counts. 
For comparison, we can measure no excess  
($\sim -30 \pm 21$ and  $\sim -6\pm 29$
counts) in the
N and S directions along the major axis in the same area and at the same 
radial distance from center.

Asymmetries and irregularities in the emission are also found on smaller scales.
Outside of the nuclear area, an almost X-shaped emission (the arm to
the SE of the major axis is not as evident as the others) is detected, as
indicated rather irregular azimuthal surface brightness distribution
outside of the nuclear area shown in Fig.~\ref{azim}.  In particular,
there are two enhancements relative to neighboring sectors at $\sim
\pm 45^\circ$ from the major axis, and clear depressions in the
direction almost perpendicular to the major axis ($\sim 13^\circ$ and
270$^\circ$ in Fig.~\ref{azim}). 

\begin{figure*}
  \resizebox{12cm}{!}{\psfig{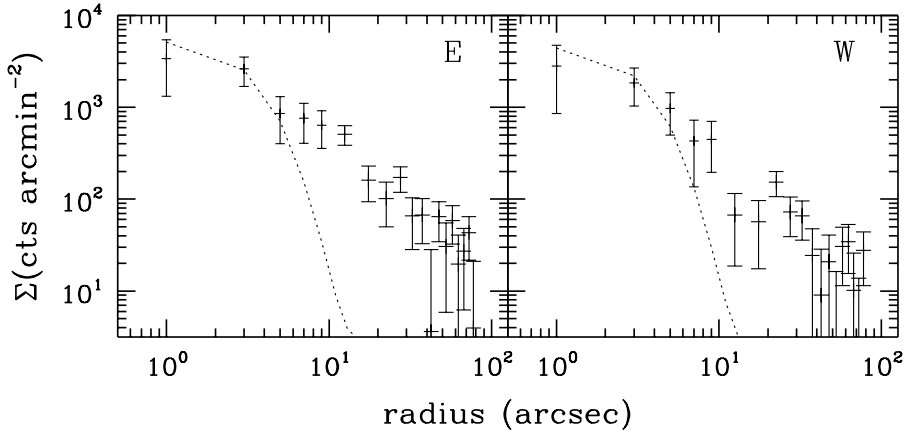}}
  \hfill
  \parbox[b]{55mm}{
    \caption[]{
     Radial profiles of the HRI  X-ray surface brightness
     of NGC~3079 in two opposite halves. The dotted line indicates the 
     profile expected for a point source arbitrarily normalized to
     the $+1\sigma$ value of the innermost point}
  \label{ew}}
\end{figure*}

To better study the presence of an unresolved core in the data,
we have further analyzed the HRI data, and we have produced radial
profiles of the net emission in two  opposite halves, $i.e.$ East and West. 
As shown by Fig.~\ref{ew}, where the comparison with the PSF 
is also shown, there is a suggestion for the presence of an unresolved source
embedded in a more extended component, and a significantly steeper decline  
in the Western half of the plane than the Eastern half.
However, the point-source component does not dominate in the inner
20$''-30''$ radius region, so any attempt to study it at the PSPC
resolution is hampered by the presence of the extended component.

\subsubsection{Spectral analysis}
\label{spectra-anal}
The morphology of the emission from NGC~3079, coupled with its optical
and \HI\ properties, suggests that the X-ray emission comes from three
separate regions with presumably very different characteristics.  The
bright central region (60\arcsec\ diameter), 
resolved in a complex source plus 2 point-like sources by the HRI, is likely to
experience a large absorption.    Similarly, the emission from the
disk, seen edge-on, 
will be heavily absorbed, with the exception of the very external
layers.  In the halo region, instead, absorption consistent with the
line of sight \HI\ column density from our own galaxy is expected.  A
possible difference between the two sides of the plane, as the galaxy is
not perfectly edge-on, might also be expected. In addition, inter-galaxy gas 
within the group (see introduction) may add additional absorption.

Given the limited statistics offered by the PSPC data, we have tried to
minimize the number of separate regions from which to extract the photons
for spectral analysis purposes.  We have
therefore checked with the aid of the hardness ratios whether different
regions showed significantly different spectral characteristics, as
could be expected from the considerations above. HR1 and HR2 have been
calculated as defined in the previous section for 4
different regions: 
\begin{description}
\item{ a)} the central region, defined as a circle of 30\arcsec\ radius,
centered at $\alpha$=$10^{\rm h}01^{\rm m}58\fs3$ and 
$\delta$=55\degr40\arcmin53\arcsec\ (J2000.0)
\item{ b)} the disk region, 
defined as a box of size $73\arcsec\times260\arcsec$,
positioned along the major axis of the galaxy, shifted 20\arcsec\ to S.  
The central source region has been masked out in the count extraction
(cut radius 30\arcsec)
\item{ c)} the halo region above the plane, defined as a box of size $3\farcm2 
\times 2\farcm0$, positioned parallel and adjacent to the E of the disk region 
\item{ d)} the halo region below the plane,  defined as a box of size $4\farcm3 
\times  1\farcm5$, positioned parallel and adjacent to the W of the disk region
\end{description}  

The background was taken from an annulus of 4\arcmin\ and 6\arcmin\ 
inner and outer radii 
respectively, centered at the X-ray peak position (source P16). 

   \begin{table*}
      \caption{Extracted spectra. Source plus background region 
       (and covered area) as well as background region are given}
         \label{spec-tab}
         \begin{flushleft}
         \begin{tabular}{llrlrrr}
            \hline
            \noalign{\smallskip}
Spectrum & Source region & Area & Background region  & Net counts   
       & HR1               & HR2         \\
            &                      &(arcmin$^2$)         & & & & \\
            \noalign{\smallskip}
            \hline
            \noalign{\smallskip}
{\it all} & 0 to 5\farcm5 diameter around&23.76 &8$'$ to 12$'$ diameter after
                                                      &1142$\pm$57.4 
       & 0.49$\pm$0.06 & 0.03$\pm$0.04   \\
          & the center of the galaxy     &      & screening point sources 
& & & \\
          &                         &      & with a cut diam. of 60$''$ 
 & & & \\
            \noalign{\smallskip}
            \noalign{\smallskip}
            \noalign{\smallskip}
{\it central} & 0 to 60$''$ diameter around& 0.79&see {\it all} & 401$\pm$21.7
& 0.79$\pm$0.05 & 0.21$\pm$0.05 \\
            \noalign{\smallskip}
              & central X-ray emiss. peak& & & & & \\
            \noalign{\smallskip}
            \noalign{\smallskip}
            \noalign{\smallskip}
{\it clocal} & see {\it central} & 0.79 & 60$''$ to 90$''$ diam. around
& 290$\pm$24.1&0.86$\pm$0.07 &0.33$\pm$0.08 \\
            \noalign{\smallskip}
&  & &central X-ray emiss. peak & & & \\
            \noalign{\smallskip}
            \noalign{\smallskip}
            \noalign{\smallskip}
{\it disk} & box from 0\farcm6 E to 0\farcm6 W & 4.45 & see {\it all} &
286$\pm$26.4 &0.77$\pm$0.12 &$-$0.08$\pm$0.07  \\ 
            \noalign{\smallskip}
           & along minor axis, from & & & & & \\
            \noalign{\smallskip}
           &2\farcm0 N to 2\farcm3 S along major & & & & & \\
            \noalign{\smallskip}
           &axis, {\it central} excluded& & & & & \\
            \noalign{\smallskip}
            \noalign{\smallskip}
            \noalign{\smallskip}
{\it halo} & 2 boxes:  0\farcm6 E to 2\farcm6 E;  &12.83& see {\it all} &
349$\pm$38.8 &0.08$\pm$0.12 &$-$0.17$\pm$0.11 \\
            \noalign{\smallskip}
           & 0\farcm9 N to 2\farcm3 S and 0\farcm6 W 
& & & & & \\
            \noalign{\smallskip}
           &to 2\farcm1 W; 2\farcm0 N to 2\farcm3 S& & & & & \\
            \noalign{\smallskip}
            \hline
          \end{tabular}
         \end{flushleft}
   \end{table*}

   \begin{table*}
         \caption{Spectral investigations of the extracted spectral files}
         \label{spec}
         \begin{flushleft}
         \begin{tabular}{llrrrrrr}
            \hline
            \noalign{\smallskip}
Spectrum& Model$^{~\ast}$
& $N_{\rm H}^{~\dag}$ & Index & $T$ &$\chi^2/\nu$
& $f_{\rm x}^\S$
       & $L_{\rm x}^\S$   \\
& & & &(keV) & & &\\
            \noalign{\smallskip}
            \hline
            \noalign{\smallskip}
{\it all} &POWL& $8_{-1}^{+2}$ 
& $3.2_{-0.3}^{+0.3}$ & & 8.1/5 &8.37 &29.9 \\
\noalign{\smallskip}
(S/N$\gid$10)  &THBR& $5_{-1}^{+1}$ & & $0.6_{-0.1}^{+1.1}$& 7.2/5 
&7.54&26.9 \\
\noalign{\smallskip}
           &THPL& $0.8^{\rm fix\, \sharp}$ & & $1.2_{-0.1}^{+0.1}$ 
&59.4/6 &5.68&20.3 \\
            \noalign{\smallskip}
            \noalign{\smallskip}
            \noalign{\smallskip}
{\it central} &POWL& $22_{-10}^{+23}$ 
& $4_{-1}^{+2}$ & & 2.1/3 &3.46 &12.4 \\
\noalign{\smallskip}
(S/N$\gid$8) &THBR& $10_{-3}^{+9}$ & & $0.7_{-0.2}^{+0.3}$ 
& 1.8/3 &2.71 &9.7 \\
\noalign{\smallskip}
           &THPL& $2_{+1}^{-1}$ & 
& $1.3_{-0.1}^{+0.2}$ & 10.8/3 &2.21 &7.9 \\
            \noalign{\smallskip}
            \noalign{\smallskip}
            \noalign{\smallskip}
{\it clocal} &POWL& $39^{+49}_{-22}$ 
& $4.2_{-1.5}^{+2.4}$ & & 0.08/4 &1.86 &6.6 \\
\noalign{\smallskip}
(S/N$\gid$6)       &THBR& $20_{-11}^{+46}$ & & $0.6_{-0.3}^{+0.6}$ 
& 0.10/4 &1.86 &6.6 \\
\noalign{\smallskip}
           &THPL& $4_{-2}^{+4}$ & & $1.5^{+0.8}_{-0.2}$ & 3.9/4
&1.74 &6.2 \\
            \noalign{\smallskip}
            \noalign{\smallskip}
            \noalign{\smallskip}
{\it disk} &POWL& $16_{-6}^{+24}$ & $4.0_{-0.8}^{+1.6}$ & 
& 2.1/4 &2.19 &7.8 \\
\noalign{\smallskip}
 (S/N$\gid$6)          &THBR& $9_{-3}^{+8}$ & & $0.4_{-0.1}^{+0.2}$ 
& 3.2/4 &1.98 &7.1 \\
\noalign{\smallskip}
           &THPL& $0.8^{\rm fix\,\sharp}$ & & $0.9_{-0.1}^{+0.1}$ 
& 19.9/5                &1.36 &4.7 \\
            \noalign{\smallskip}
            \noalign{\smallskip}
            \noalign{\smallskip}
{\it halo} &POWL& $8^{+4}_{-3}$ & $3.9_{-0.6}^{+0.7}$ 
& & 11.1/3 &2.38 &8.5 \\
\noalign{\smallskip}
(S/N$\gid$6)           &THBR& $4_{-1}^{+2}$ & 
& $0.4^{+0.1}_{-0.1}$ & 6.1/3 &2.14 &7.6 \\
\noalign{\smallskip}
           &THPL& $0.8^{\rm fix\,\sharp}$ & & $0.3_{-0.1}^{+0.1}$ 
& 10.5/4 &1.72 &5.8 \\
            \noalign{\smallskip}
            \hline
          \end{tabular}
         \end{flushleft}
\[
\begin{array}{lp{0.95\linewidth}}
$$^{\rm \ast}$$ & POWL: power law, THBR: thermal bremsstrahlung,
THPL: thin thermal plasma\\
$$^{\dag}$$ & in units of 10$^{20}$ cm$^{-2}$ \\
$$^{\rm \S}$$ & in units of 10$^{-13}$\,\ergcm\  
and 10$^{39}$\,\ergsec, respectively, for
0.1--2.4~keV band, corrected for Galactic absorption
and calculation of 1$\sigma$ errors\\
$$^{\rm \sharp}$$ & fixed to Galactic foreground for spectral fits \\
\end{array}
\]
   \end{table*}

We found no significant difference between the hardness ratio values for
regions c) and d), and we have therefore defined as $halo$ region the
combination of c)+d) above.  Table~\ref{spec-tab}
summarizes the definitions of source and background regions used to determine
hardness ratios and photon energy distributions for spectral fitting.
As shown by Table~\ref{spec-tab}, we have also considered the galaxy as
a whole, and we have used a local background for the central region, to
take into account possible contamination from the disk.

Comparison of the HR1 and HR2 values and also  with the $theoretical$
hardness ratios shown in the plots of
Fig.~\ref{hr-theor} clearly indicate that the central region, the disk and the
halo occupy different regions of the HR1/HR2 diagram.
We have then used simple spectral models to fit the data in the different
regions, as indicated in Table~\ref{spec}. Raw spectra have been rebinned
to obtain at least the signal to noise level per bin given in col. 1
of Table~\ref{spec}. Rough errors for the fluxes and luminosities in
Table~\ref{spec} are indicated by the statistical errors on the 
net counts (see Table~\ref{spec-tab}, although additional
uncertainties come from the poor knowledge of the spectrum, as
can be seen by simply comparing the fluxes derived for different 
models).  

It is apparent that in all cases but the halo region, power law and
thermal brems\-strah\-lung models give a better approximation of the
data than the thin plasma model (Raymond-Smith code).  Moreover, this
latter would prefer a low energy absorption $below$ what is expected
from the line-of-sight \HI\ column density, without giving a
significant improvement in the best fit $\chi^2$ value.  Power law or
thermal brems\-strah\-lung models give essentially equivalent goodness
of fit, and produce spectral models that are a good approximation of
the energy distribution of the detected photons.  In all cases a
significant amount of intrinsic absorption above the line-of-sight
value of $8 \times 10^{19}$\,cm$^{-2}$ is suggested, consistent with the
idea that the emission comes from within the galaxy, and therefore
suffers from the absorption in NGC~3079 itself.   The fact that we have
obtained very  similar results from the disk and central regions is not
surprising since, as already remarked above, the extended emission
contributes significantly even at small radii, and this, combined with
the extremely poor statistical significance of the data, does not allow
us to distinguish the presence of a different component (for example
from the point source in the nuclear region suggested by the HRI
data).

Even though a good fit is already obtained for power law and thermal
brems\-strah\-lung models, 
for the galaxy as a whole we have also tried to improve on the best
fit values in the thin thermal plasma model by assuming two
temperatures.   This is done mostly to compare ours with published
results on similar objects, and it is partly
justified by the fact that the requirement of a lower-than-galactic 
absorbing column in the 1-temperature fit could be suggestive of an 
additional very soft component.  Indeed we find best fit values of 
kT$_1 \sim 0.3-0.5$ and kT$_2 \ge 1$ keV, for a the minimum $\chi^2$ 
reduced to an acceptable value of 5.3 (for 4 degrees of freedom) 
and the best fit N$_H$
consistent with the Galactic value.

The halo region cannot be fit by any of the simple models ($i.e.$ the
minimum $\chi^2_\nu$ value is larger than 2 in all cases).  
In spite of the limited significance of our procedure, we have
nevertheless tried
to fit the data with both a two-component model ($i.e.$ two
brems\-strah\-lung models and a brems\-strah\-lung and a thin
plasma model) and with a
thin plasma model with varying abundances.  In both cases the minimum
$\chi^2$ value reduces drastically to perfectly acceptable values 
(1.5 and 2.2) and the best fit values are kT 0.2 and 1 keV (2-T model)
and kT 0.5 keV, 5\% solar abundance.  
While it is
therefore possible that more sophisticated models might be required for
this region, given the limited statistical significance of the data,  we
cannot discriminate between different scenarios, nor can we be sure
that our more sophisticated modeling of the data is correct, since we
are left with 1 or 2 degrees of freedom.  Since the resulting fluxes
that we can derive with the different best fit models are all very
similar (see Table \ref{spec}), 
we will therefore assume the best fit values from
the thin thermal plasma model for counts to flux conversion purposes for
the halo emission.


%
%
%

\subsection{NGC~3079 nuclear X-ray emission}
\label{results_nucleus}
While in the PSPC the nuclear source can not be resolved (see 
Fig.~\ref{four_in_one}), a complex source is resolved with the HRI
resolution at the galaxy's center (Fig.~\ref{master_hri_center}) with an
extent of the order of $20\arcsec \times 30\arcsec$\,($1.7 {\rm kpc} \times 
2.5$\,kpc).  
In addition, a connected peak
at $\sim 25\arcsec$ distance north of the nucleus and a separate peak 
to the SW (source H13) can be clearly seen.  

The radial distribution of the emission, centered on the X-ray peak, and
its morphology both indicate that the source is extended and structured
and that a  possible point source located at the nuclear
position could only contribute $\sim 1/3$ of the emission in the area.  
If this source indeed coincides with the active nucleus of the galaxy,
its estimated count rate of $\sim 2 \times 10^{-3}$\,cts s$^{-1}$ would
correspond to a luminosity L$_{\rm x} > 5 \times 10^{40}$\,\ergsec,
calculated assuming a thermal brems\-strah\-lung model with a low energy
absorption equivalent to a column density of $\sim 5 \times 10^{20}$\,
cm$^{-2}$.  The lower limit sign is due to the fact that the absorption
in the nucleus is likely to be much higher than the value assumed
(which corresponds  to an average column density in the disk on NGC~3079, see
\HI\ maps (Irwin \& Seaquist 1991) 
and best fit parameters from spectral fits).
While a different choice of the spectral model would give very similar
values (see Table~\ref{spec}), the low energy absorption adopted influences
very strongly the estimate of the intrinsic flux in the ROSAT band.
mm-wave estimates of the extinction towards the nucleus of this source
indicate that a minimum of $1.4 \times 10^{24}$ H$_2$ cm$^{-2}$ 
should be expected (Sofue \& Irwin 1992). 
ASCA data (Serlemitsos et al. 1997; Dahlem et al. 1998) 
in fact suggest the presence of a hard and heavily absorbed 
component (with absorbing column in excess of $10^{22}$
cm$^{-2}$; however notice that Ptak et al. (1998) give much lower 
best fit values for absorption to the power law component)
in the spectrum
of NGC 3079 as a whole (the spatial resolution of ASCA allows only a
global measure of the spectral properties of NGC~3079), with a
luminosity L$_{\rm x}$(2-10 keV) of $2 \times 
10^{40}$\,\ergsec.   The origin of this emission is not as yet
unambiguously interpreted, since it is present both in galaxies with
recognized nuclear activity (Low luminosity AGN, LINER) and in 
starburst galaxies 
(Serlemitsos et al. 1997; Ptak et al. 1998; Dahlem et al.
1998); therefore, due to the lack of good spatial resolution at
high energies, it can either be related to the nuclear activity or to the 
binary population and starburst phenomenon (or both).   
Given the high absorption suggested by the data and the presence of
nuclear activity, in NGC~3079 this component could come from a 
very absorbed compact nuclear source.  In this case  
we expect about $6 \times 10^{-4}$\,cts s$^{-1}$ in the nucleus 
with the HRI. Given the large uncertainty (also in absorption) 
this rate could be consistent with the possible point source
contribution from a nuclear source in the HRI image (see above).
  
Given the strong contamination from the diffuse background, and the
small statistics (about 40 cts), 
we cannot really measure possible time variability in the
flux from this source, which would confirm its point-source nature
and propose an identification with a super-luminous X-ray binary close to
the nucleus or the X-ray detection of the NGC~3079 active nucleus itself.

%
%

\subsection{NGC~3073 and MCG~9-17-9}
\label{companion_results}
We searched for X-ray emission from the companion galaxies of NGC~3079, 
NGC~3073 and MCG~9-17-9. Only MCG\ 9-17-9 was detected  
(H6/P9).   The 69 counts  detected with the PSPC are insufficient for a
detailed spectral fitting, and we could only determine one of the two
hardness ratios (see Table~\ref{sources_pspc}).  
Comparison with the plots of Fig.~\ref{hr-theor} indicates that for these
two spectral models the
source is probably absorbed above the line-of-sight value, and that its
spectrum is relatively hard ($i.e.$ kT $>$0.5 keV or $\alpha <1$).  

A weak enhancement is seen at the position of NGC\ 3073 both in the HRI and
in the PSPC, however with a significance far below the threshold for
our source catalogs.  

\begin{table}
\caption{X-ray parameters for NGC~3073 and MCG~9-17-9 
}
\label{companions}
{
\begin{flushleft}
\begin{tabular}{llrrr}
\hline
\noalign{\smallskip}
Galaxy & &Count rate & f$_{\rm X}$ & L$_{\rm X}$ \\
\noalign{\smallskip}
       & & (10$^{-4}$s$^{-1}$) &$\ast$ & $\ast\ast$ \\
\noalign{\smallskip}
\hline
\noalign{\smallskip}
NGC~3073   &PSPC & 0.8$\pm$0.5 & 0.9$\pm$0.6 & 0.3$\pm$0.2 \\
           &HRI  & 0.3$\pm$0.2 & 1.3$\pm$0.9 & 0.5$\pm$0.3 \\
\noalign{\smallskip}
MCG~9-17-9 &P9 & 38$\pm$\phantom{0.}6 & 44$\pm$\phantom{0.}7 & 
15.7$\pm$2.5 \\
           &H6 & 8.4$\pm$2.2 & 33$\pm$\phantom{0.}9 & 11.8$\pm$3.2 \\
\noalign{\smallskip}
\hline
\end{tabular}
\end{flushleft}
\[
\begin{array}{lp{0.95\linewidth}}
$$^\ast$$ & fluxes in units of 
10$^{-15}$\,\ergcm\
for a 5~keV thermal bremsstrahlung spectrum in the 0.1--2.4~keV
band, corrected for Galactic absorption\\
$$^{\ast\ast}$$ &luminosity in units of 10$^{38}$\,\ergsec\  
assuming a distance of 17.3 Mpc 
\end{array}
\]
}
\end{table}

Table~\ref{companions} summarizes PSPC and HRI count 
rates, X-ray fluxes f$_{\rm X}$ and luminosity L$_{\rm X}$ for these
sources.  To convert count rates to fluxes we assumed a 5 keV thermal 
brems\-strah\-lung spectrum (see Table~\ref{conversion});
the same distance as for NGC~3079 was assumed for 
the luminosity determination.   

The comparison of the HRI and PSPC flux of
MCG~9-17-9 indicates that the source has not varied between the two
observations.  
 
\section{Discussion}
\begin{figure*}
  \resizebox{12cm}{!}{\psfig{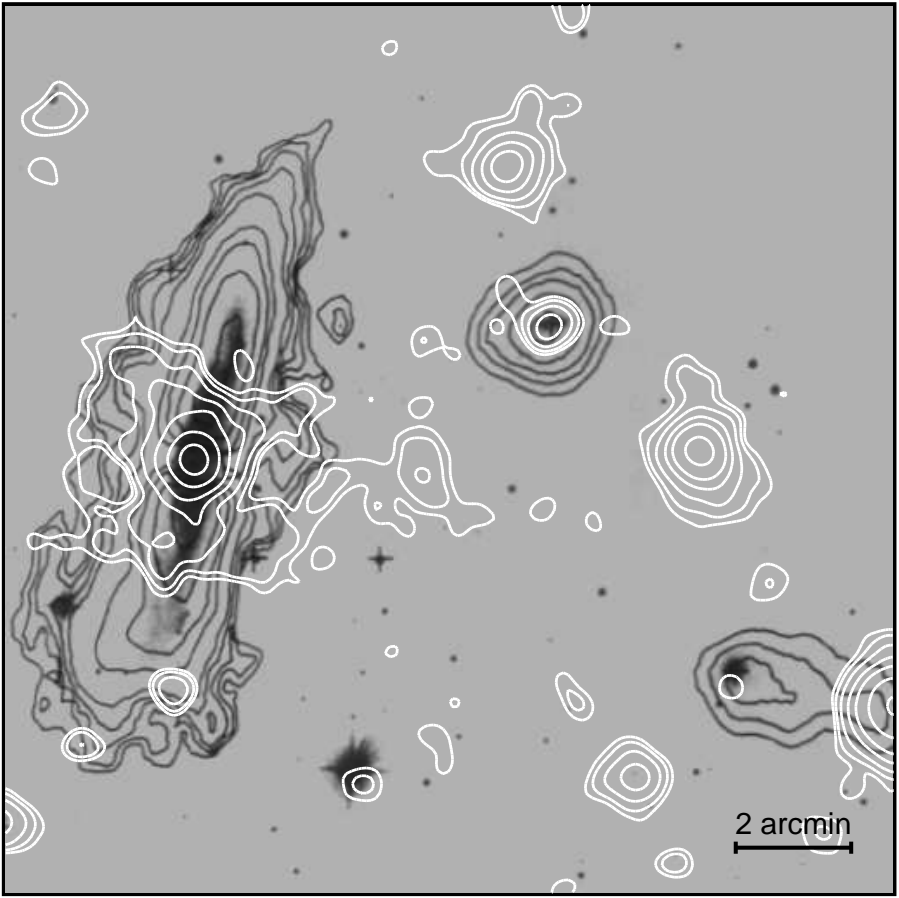}}
  \hfill
  \parbox[b]{55mm}{
    \caption{
     Contour plot of the broad band PSPC X-ray emission of the region of
     NGC~3079 and its companions
     overlaid in white on Fig. 1 of Irwin et al. 1987,
     showing \HI-contours in black superposed on an reproduction of the
     POSS red plate.
     The same contour levels as in Fig.~\ref{image_pspc} are used}
    \label{opt_hi_pspc}}
\end{figure*}
The ROSAT PSPC observations of the edge-on NGC 3079 have shown a
complex emission in and outside of the plane of the galaxy.  The global
X-ray luminosity in the ROSAT band is $\sim 3 \times 10^{40}$\,\ergsec\ 
(d=17.3 Mpc),
higher ($\sim$ factor 10) than
what is observed in other edge-on late-type galaxies of similar optical 
luminosity (for example in NGC 4631,
Vogler and Pietsch 1996; NGC 4565, NGC 5907, Vogler et al. 1995). 
The emission has been separated into three components of similar brightness 
- disk, central region, and halo - that will be discussed
separately in the following subsections. 

Two explanations for the enhanced X-ray luminosity of NGC 3079 can be 
put forward that are triggered by anomalies of the galaxy measured 
in other wavelength regimes: 
\begin{enumerate}
\item The galaxy belongs to a group of  powerful far infrared emitters,
with a far infrared excess comparable to that of starburst galaxies
like M82, NGC 253, and NGC 2146 (Lehnert \& Heckman 1995).    In fact
the X-ray, optical and far infrared luminosities of NGC 3079 and NGC
2146 are almost identical (Armus et al. 1995). As we discuss later
similarities of the galaxies do not stop there.  It is
therefore likely that enhanced star formation activity is not only present
in the nuclear area but rather
widespread in the disk, and therefore it is not remarkable that the
X-ray luminosity is also enhanced.  
\item On the other hand NGC 3079 harvests a low
ionization nuclear emission line region (LINER) or Seyfert 2 type nucleus
(Heckman 1980, Ford et al. 1986). In addition, 
it is known to contain a bright continuum nuclear radio source
and bipolar jet like structures emerging from the nucleus along the projected 
minor axis (extent 50\arcsec\, see Fig.~\ref{halpha_hri}). Similar features and
classification of the nucleus also apply for NGC 4258 and M51
and even for the starburst galaxy M82, Tsuru et al. (1997) argue from the 
analysis of wide-band X-ray spectra for the presence of an obscured 
low-luminosity AGN. These galaxies 
were detected as just as bright extended X-ray emitters by ROSAT 
(e.g. Pietsch et al. 1994, Ehle et al. 1995, Read et al. 1997). 
This makes it conceivable that jet-like outflows from  
active nuclei might indicate enhanced extended X-ray emission.  
\end{enumerate}

While it is not clear which of these types of activity (or even both?)
is responsible for the enhanced X-ray emission,
both may have been triggered by galaxy galaxy encounters in the NGC~3079 
galaxy group that also contains NGC 3073 and MCG 9-17-9
(see Sect.~\ref{group}). 

The low X-ray flux associated with NGC 3079 
has not allowed us to properly measure the spectral characteristics of 
the emission components, therefore a physical interpretation of 
the results is not possible. As shown in Sect. 3.3.2,
a power law model with relatively steep spectral index or a soft thermal
brems\-strah\-lung could fit the data equally well.  However, the power law
index is in all cases very steep ($\alpha_E \ge 2$) and also the temperature
of the thermal bremsstrahlung spectra is rather low if we compare them
for instance with spectral fits to individual X-ray binaries or the
integral bulge spectrum (primarily X-ray binaries) found with the ROSAT 
PSPC for M31 ($\alpha_E\ 0.3 - 0.6$ or temperatures above $\sim 1$ keV, 
Supper et al. 1997).  It is likely that the simple models that we have
assumed here for lack of statistics are inadequate to represent the more
complex characteristics of the X-ray emission in these objects 
(for example a population of individual sources plus a multitemperature or 
non-equilibrium interstellar medium). 
Dahlem et al. (1998) have shown that a correct interpretation of the
spectra of galaxies requires data in a larger energy
range than provided by the PSPC data alone, even though there are still
residual ambiguities in the correct model to be used.  
In particular they find in several instances that two gas phases are
needed in addition to a power law component at high energies.  The lack of
spatial resolution however does not allow a proper
investigation of the location of these individual components, 
that cannot therefore be unambiguously identified with the separate
sources of emission in galaxies.   In this respect, only future X-ray
missions, that combine high sensitivity with good spectral and spatial
resolution over a large energy range will allow a significant step
forward in our understanding of the X-ray properties of galaxies. 
From the present data however we could be tempted to assume that indeed
a multiphase interstellar medium is present and interpret 
the (very crude)  indication of the hardness ratios in terms of temperature
variations in the disk versus central region,
and assign a higher temperature to the emission coming from the
central source coincident with the super-bubble (see Sect. 4.2).  This
would be analogous to the results from the more detailed analysis of
NGC~253, for which the central starbursting region is harder than the
surrounding disk (Dahlem et al. 1998).  


In either case however
a large low energy absorption is suggested, well above the line-of-sight
\HI\ column density.  This result is not surprising since the disk
is embedded in an extended \HI\ disk (Irwin et al. 1987).  
Only at large distances
from the plane could local obscuration be neglected, and indeed the
(though extremely uncertain) results for the halo region do not require large
absorbing columns. 



\subsection{X-ray emission from the disk of NGC 3079}
Several bright sources are detected in the galaxy plane of NGC 3079
(Table~\ref{n3079_sources}), with
luminosities higher than 10$^{38}$\,\ergsec, $i.e.$ well above the
Eddington luminosity for a 1 M$_\odot$ accreting object.  As already
discussed, these could be real point-like sources or could be local
enhancements in a more diffuse emission.   However,  the optical image of the
galaxy shows a very patchy appearance, and there are a lot of giant
\HII\ regions in the plane (Veilleux et al. 1995).  Over-luminous
\HII\ regions, relative to those observed in our galaxy and
in other normal spirals like M101, for which L$_{\rm x} \le 10^{38}$\,\ergsec\
(Williams \& Chu 1995), have already been observed in
relation with the presence of enhanced star formation ($e.g.$ in the
Antennae, Read et al. 1995, Fabbiano et al. 1997). 
As discussed above, the integral spectrum of the disk region is rather 
soft compared to bright X-ray binary spectra and can be taken to favor 
the over-luminous \HII\ region origin of the emission. 
Deep observations with improved spatial and spectral resolution are needed
to solve the origin of the disk emission.

%

\subsection{X-ray emission from nuclear super-bubble of NGC 3079}
NGC 3079 hosts the most powerful example of a wind-blown bubble to the east 
of its nucleus.  Recent spectroscopic studies in the near infrared 
and optical however argue against a starburst origin for the powering
of the 
wind, and suggest an active nucleus with somewhat exceptional properties
(Hawarden et al. 1995).  In this case, one would expect a wind from an
active galactic nucleus (AGN, see Heckman et al. 1986) -- 
how can X-ray observations help understanding the nature of this source?

\begin{figure*}
 \resizebox{12cm}{!}{\psfig{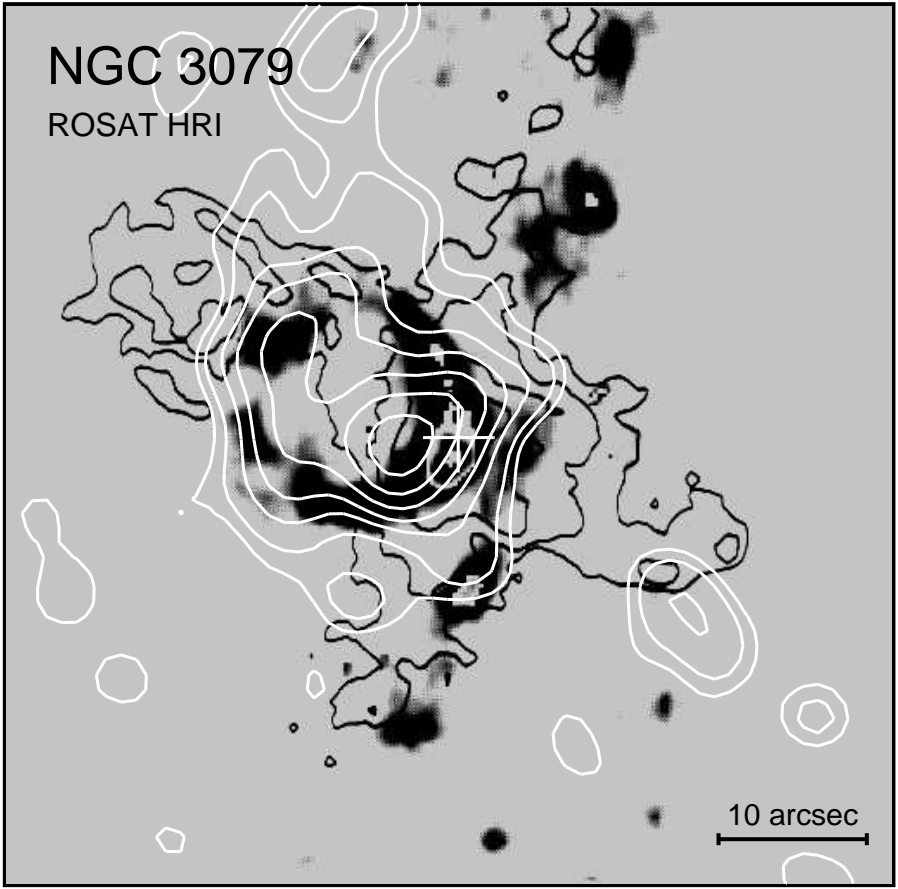}}
 \hfill
 \parbox[b]{55mm}{
  \caption[]{
   Contour plot of the central emission region of
   NGC~3079 for ROSAT HRI overlaid on Fig. 1 of
   Veilleux et al. (1994), showing the 
   continuum-subtracted distribution of H$_{\alpha}$
   + [N\,{\sc ii} $\lambda\lambda$6548, 6583 line
   emission (grey-scale) and the 20 cm continuum
   distribution (black contours).
   X-ray contours are given in white (same levels 
   used as in Fig.~\ref{master_hri_center}) }
   \label{halpha_hri}}
\end{figure*}

Emission in the center of NGC~3079, coincident with the H$\alpha$ loop
(Ford et al. 1986)
and the kiloparsec scale radio lobes (de Bruyn 1977, Duric \& Seaquist
1988) is clearly detected in the HRI data.  The source appears complex
(see Fig.~\ref{halpha_hri}).  A possible unresolved component, apparently
shifted by about 3\arcsec\ to the east from 
the position of the optical nucleus (marked by the cross in 
Fig.~\ref{halpha_hri}) could be --  within the systematic position errors -- 
coincident with the nucleus. This
source, if identified with the hard spectral 
component measured by ASCA (see discussion in Sect. 3.4) that has been
associated with the emission of an highly absorbed AGN, is brighter by a 
factor of 3 -- 4 than the extrapolation of the ASCA spectrum 
into the ROSAT band would suggest.
If the positional displacement however is real, it could represent the 
brightest and hottest part of the out-flowing material that is not shielded
by the absorbing material in the central disk. Such a displacement to the 
east is expected in such a scenario due to the high absorption in the
inner disk of NGC~3079 and the disk orientation. The inclination of 84\degr\
and the orientation of the galaxy (the western side of the
disk is closer to us than the eastern side) lead to reduced absorption in the
direction to the nuclear area in the east and high absorption in the west 
(N$_{\rm H}$ in the inner disk is $> 10^{22}$ cm$^{-2}$ according to 
Irwin \& Seaquist 1991). The high absorption may also cause that no 
X-ray or H$\alpha$ emission coinciding with the western component of the 
radio jet is detected.
 
A more extended component elongated to the
north is possibly connected to a feature seen at
larger scale (Fig.~\ref{halpha_hri}), and may 
indicate a preferred channel to fuel the emission in
the halo.    

The HRI attitude solution carefully derived in Sect. 2.1 clearly contradicts
the solution put forward by Dahlem et al. (1998) which is shifted to the
north by 15\arcsec\ and -- if correct -- would exclude a connection of the 
central X-ray emission with either the nucleus or the nuclear super-bubble 
of NGC 3079. 

The small extension of the bright central X-ray emission 
region compared to the PSPC spatial
resolution coupled with it's low flux prevents us from reliably determining 
spectral properties.  In the innermost 30$''$
radius, that correspond to the minimum cell size for count extraction
that is reasonable for PSPC data, the HRI resolution indicates that at
least three different components are present, all contributing to the
overall emission: the disk with two non nuclear point-like sources, 
the nuclear bubble and the unresolved point
source.  Neither of these dominates, with the possible exception of the
disk emission, which can in part be subtracted by choosing a local
background (see Table~\ref{spec-tab}), and therefore a spectral analysis should
take all of them into account.   This is clearly unrealistic given the
small number of photons that can be collected in the area and also given
the limited spectral capabilities of the PSPC, that indeed indicate a
single model as acceptable.  Whether one can however use the results of
any such fit and interpret them as giving a physical interpretation of
the observations remains questionable. Observations with better
statistics and higher spectral and spatial resolution and an extension
to higher energies are needed badly. Only then will we be able to answer
the question if the -- with ROSAT HRI -- unresolved component 3\arcsec\
to the east of the optical nucleus is coincident with the optical nucleus 
and the hard ASCA source (see above) or
-- our prefered explanation -- the brightest and hottest part of the 
out-flowing material. 

We can compare the emission from the nuclear super-bubble to the X-ray 
emission from the plume
detected in NGC~253 (see Pietsch et al. 1998), the anomalous arms of NGC~4258
(see Pietsch et al. 1994), and the extended
emission resolved with the HRI in NGC~2146 (Armus et al. 1995),
X-ray features of similar appearance and proposed origin. While in 
the prototypical starburst galaxy NGC~253 the plume has an extent 
perpendicular to the major axis of $\sim 700$ pc and a luminosity of 
$2 \times 10^{39}$ \ergsec, in the LINER galaxy NGC~4258 the extent is 
1.5 kpc and the luminosity is $5 \times 10^{39}$ \ergsec. For the starburst
galaxy NGC~2146 (after correcting for an inclination of 36\degr, Tully (1988))
the HRI extent is 2.2 kpc and the unresolved luminosity in the range of
$5 \times 10^{39}$ \ergsec. Due to the low inclination of NGC~2146, however,
it is not possible to determine the scale height of the extended emission
perpendicular to the galaxy disk and separate point-like and extended
components. In comparison
the NGC~3079 central component has an extent of 1.7 kpc and a luminosity
of $7 \times 10^{39}$ \ergsec. NGC~253 and NGC~2146 show coinciding
optical line emission, that is explained by an out-flowing wind driven by 
the nuclear starburst. Close to the nucleus of NGC~253 embedded in the plume 
emission a slightly extended source represents the hottest
part of the out-flowing material and may reflect the point-like nuclear
component in NGC~3079.
The extended nuclear features in NGC~3079 and NGC~4258 
are more X-ray luminous than in the pure starburst galaxies NGC~253 and 
NGC~2146 and are not only traced by optical emission lines 
but in addition by jet-like 
continuum radio emission indicating the importance of magnetic fields 
funneling the outflow and the existence of a driving AGN.


\subsection{Extended X-ray halo of NGC~3079}
The PSPC contour plots (Figs. \ref{image_pspc} and \ref{opt_hi_pspc}) 
and the radial distribution of the PSPC and HRI detected photons 
(Fig. \ref{radplot}) clearly indicate that the emission from NGC~3079
extends to a radius of 13.5 kpc. The distributions along the major and 
minor axis (Fig. \ref{rad_maj_min}) demonstrate that this emission is not
only originating within the NGC~3079 disk but is filling the halo to at least
the inclination corrected D$_{25}$ diameter of the galaxy. The spectral
investigations clarified that the halo emission contributes about one third 
to the total X-ray luminosity ($\sim 2.7 \times 10^{40}$\ergsec) of the
galaxy in the 0.1 -- 2.4 keV ROSAT band; emission from the center and the disk
complete the luminosity in about equal parts. 

The extent of the halo of NGC~3079 
is comparable or slightly bigger than the one of the 
starburst galaxy NGC~253 (Pietsch et al. 1996, 1998) and bigger 
by a factor of $\sim 4$ than that of
the active galaxy NGC~4258. On the other hand, the
luminosities (and temperatures) for the halos are similar in the 
two galaxies with known active nuclei,while they are higher than
in the starburst galaxy. The halo emission
has a X-shaped structure (c.f. Figs. \ref{azim} and \ref{opt_hi_pspc})
similar to NGC~253 (however not as
clearly resolved due to the greater distance of NGC~3079).  
X-shaped filaments of the diffuse ionized medium within a radius of 5 kpc, 
emerging from the inner disk and rising more than 4 kpc above the disk plane
of NGC~3079, have been reported from optical {\hbox{[N\,{\sc ii}]}} 
measurements (Heckman et al. 1990, Veilleux et al. 1995).  According to 
Veilleux et al.,  the morphology, kinematics, and excitation of the
filaments suggests that they form a biconic interface between the undisturbed
disk gas, and gas entrained in a wide-angle outflow. In \HI\ observations of
NGC~3079 numerous arcs and filaments are present 
extending away from the plane of the edge-on galaxy
which seem to be unrelated to the nuclear activity (Irwin \& Seaquist 1990).
They may indicate gas flow between disk and halo in addition to  
violent outflow from the active nucleus (Heckman et al. 1990, Filippenko \&
Sargent 1992). 
 
If we assume that the halo X-ray emission is due to hot gas with a temperature
of $3.5 \times 10^6 $K (cooling coefficient according to Raymond et al. (1976)
of $5.5 \times 10^{-23} $erg cm$^3$ s$^{-1}$), 
a luminosity of $5.8 \times 10^{39} $\ergsec 
(see Table \ref{spec}) distributed in a sphere of 13.5 kpc radius with
filling factor $\eta$ we can calculate parameters of the plasma
using the model of thermal cooling and ionization equilibrium of Nulsen et al.
(1984). The electron densities, masses, and cooling times of the X-ray 
emitting gas in the halo are: 
$n_e = 7 \times 10^{-4} \times \eta^{-0.5}$ cm$^{-3}$, 
$M_{gas} = 2.0 \times 10^8 \times \eta^{0.5}$ M$_{\sun}$,
$\tau = 1.5 \times 10^9 \times \eta^{0.5}$ yr.

These gas parameters differ significantly from the parameters derived by
Read et al. (1997) for NGC~3079 which only use a strongly simplified 
geometrical model attributing the X-ray emission to a nuclear point source
and just one diffuse extended emission component. 
They find a luminosity that is higher by a
factor of $\sim 5$ and nearly twice the temperature compared to the halo 
parameters as given in 
Table \ref{spec}; {\it i.e.} their values mix up halo and disk, leading to
higher electron densities and gas masses, and slightly longer cooling times.
Read et al. analyzed several nearby spiral galaxies using similar simplified 
models. The resulting parameters for the diffuse gas always overestimate a 
possible halo component preventing a sensible comparison to the NGC~3079 halo
results.

As mentioned above to derive the gas parameters we assumed a spherical 
distribution with constant temperature. This approximation however, only
describes the situation to first order. At least two further details indicated
by the ROSAT data should be addressed:
\begin{itemize} 
\item There is a clear E -- W asymmetry  (c.f. Fig. \ref{rad_maj_min}) 
and an indication of a X-shape 
azimuthal distribution in the PSPC broad band counts 
that will be further discussed below. 
\item The PSPC contour plots of the S band emission of NGC~3079 
show far more azimuthal structure than the more spherically distributed
H1 band emission. This indicates local variations of the halo X-ray 
spectrum indicative of either changing foreground or intrinsic (within the 
halo of NGC~3079) absorption and/or temperature variations within the halo
gas. 
\end{itemize}
To better understand and characterize these effects spatially resolved spectra
are strongly needed which, due to the low counting statistics, can not be 
tackled with the ROSAT data at present. Temperature profiles of the diffuse
emission will help to solve the question of whether the E -- W asymmetry can be
attributed to inhomogeneities in the galaxy halo or instead it 
reflects a hot medium connected to the potential of the
NGC~3079 group (or both components could be present). 
This question is also relevant when interpreting \HI\ data of the companion
galaxy NGC~3073 which exhibit an elongated tail  aligned with the nucleus 
of NGC~3079; an explanation of the tail 
due to ram pressure stripping by out-flowing 
gas from NGC~3079 has been put forward in favor of stripping by movement in 
an intergroup medium (Irwin et al. 1987). A galactic 
wind perpendicular to the disk in the general direction of the minor axis 
can explain the X-shaped halo structure of the X-ray and 
optical {\hbox{[N\,{\sc ii}]}} data (see above).

\subsection{NGC~3079 as LINER/Seyfert 2 galaxy}
The unified model for active nuclei (Antonucci 1993) has been proposed to
explain the properties of the different types of AGN. 
In this scheme the difference between 
Seyfert 1 and 2 galaxies is determined by the viewing angle to the nucleus
that is embedded in a torus of molecular material.
Type 1 Seyfert galaxies allow direct observation of the nucleus, whereas for 
type 2 galaxies nuclear X-ray emission can only be seen via reflection above 
or below the torus or dust gains, or via transmission through this torus.
X-ray spectra of the nuclear emission are directly probing the absorption depth 
under which an AGN is seen, and therefore are an important test for the 
unified scheme. 

As discussed in Sects. 3.4 and 4.2, only very little direct emission from the 
LINER/Seyfert 2 type
nucleus of NGC~3079 -- if any at all -- is detected in the ROSAT band. 
Due to the many different equally bright components that contribute to the 
overall X-ray spectrum of NGC~3079 and due to limited 
counting statistics, integral spectra of the galaxy as obtained by ASCA can not 
fully separate the nuclear emission component
and only give an indication of a highly absorbed power law component
that could be related to the nucleus. 
X-ray spectroscopy with high spatial resolution as expected from the 
next generation of X-ray observatories (AXAF, XMM) will resolve the ambiguity.
From mm-wave estimates of the extinction to the nucleus of NGC~3079 
(Sofue \& Irwin 1992) and by comparison
with integral broad band X-ray spectra of other Seyfert 2 galaxies obtained
with the ASCA or BeppoSAX satellites (see Maiolino et al. 1998 and 
references therein) one may infer absorption depths well in excess of 
10$^{24}$ cm$^{-2}$ also for NGC~3079. 

Soft X-ray emission in excess of the high energy spectrum has been 
discovered in many Seyfert 2 galaxies. In Seyfert 1 galaxies such a component
may not be observed because it is covered by
the bright low absorbed flat power law spectrum.
In the nearby Seyfert 2 galaxies 
NGC~4258 and NGC~1068 the soft component could be resolved and 
is attributed to emission from 
a bipolar jet and from the halo of the galaxies (see Pietsch et al. 1994,
Wilson et al. 1992). As reported above NGC~3079 show the same components.
Therefore, a luminous X-ray halo gas heated by and
ejected via jets from the nucleus may not only explain the spatially unresolved 
soft X-ray emission components reported in other Seyfert 2 galaxies but 
also be expected in many Seyfert galaxies and AGN.    
  
\subsection{Companion galaxies NGC~3073 and MCG~9-17-9}
\label{group}
NGC~3079 (total galaxian mass $1.2 \times 10^{11}$ M$_{\sun}$) 
is the dominant galaxy in a group containing NGC~3073 
($1.1 \times 10^{9}$ M$_{\sun}$) and MCG~9-17-9 
($3.9 \times 10^{9}$ M$_{\sun}$)(Irwin et al. 1987).
While little is known from observations at other wavelength on MCG~9-17-9 -- 
it's redshift and \HI\ contours may even suggest that it is not physically 
associated with NGC~3079 -- NGC~3073 is discussed as a starburst galaxy 
situated along the trajectory of the outflow of NGC~3079 with the starburst 
induced by the super-wind (Filippenko \& Sargent 1992).
 
As shown in Sect. 3.5, MCG~9-17-9 is bright in X-rays with a luminosity 
of $\sim 1.5 \times 10^{39}$\ergsec while NGC~3073 is barely detected 
($< 10^{38}$\ergsec). The luminosity of MCG~9-17-9 is rather high for
such a dwarf galaxy (c.f. Markert \& Donahue 1985) and may indicate the
presence of at least one super-luminous source. Such a source could be a
low luminosity active nucleus, a X-ray binary radiating at super-Eddington
luminosity, a young supernova or a combination of the above. An alternative  
explanation could be the existence of several X-ray binary sources 
radiating close to the Eddington limit for a one M$_{\sun}$ star. A similarly
high luminosity has been reported for the nearby Magellanic-type star forming 
galaxy NGC~4449 (Vogler \& Pietsch 1997). There the
emission could be resolved in seven point-like sources and an additional
diffuse component of about equal luminosity.
The upper limit to the luminosity of NGC~3073 is within the values
expected for a galaxy of this mass. The starburst activity of the galaxy
does not reflect in it's X-ray luminosity. 

\section{Summary}
We have reported the results of the first detailed analysis of ROSAT data 
of the NGC~3079 field. We have paid
special emphasis on the best attitude solution to
achieve optimal HRI point spread function and positional accuracy.
With HRI and PSPC we detected 23 and 34 sources
within  35\arcmin\ diameter and a box of 35\arcmin$\times$35\arcmin, 
respectively,
apart from complex emission from the inner 5\arcmin\ around NGC~3079.
We have identified possible counterparts for several of the sources outside
NGC~3079 by comparison with optical plates and catalogues. 

The LINER/Seyfert 2 galaxy NGC~3079 manifests itself as a complex X-ray source
with a luminosity of $3\times 10^{40}$\ergsec\ and
could be resolved into three components:

(1) Extended emission in the inner 20\arcsec$\times$30\arcsec\ with a 
luminosity of $1\times 10^{40}$\ergsec can be resolved with the HRI and is 
coincident with the super-bubble seen in optical
images. The active nucleus may contribute to the emission as a point source.
We, however, prefer an explanation of this emission as the brightest and
hottest part of the out-flowing material similar to the slightly extended 
emission emission close to the galaxy nucleus within the plume of NGC~253
(Pietsch et al. 1998).     

(2) Emission from the disk of the galaxy has a luminosity of 
$7\times 10^{39}$\ergsec and 
can be partly resolved by the HRI in 3 point sources with luminosities of
$\sim 6 \times 10^{38}$\ergsec each. The soft spectrum of the disk can
be explained by a mixture of X-ray binaries and over-luminous \HII\ regions.

(3) Emission from the halo has a luminosity of $6\times 10^{39}$\ergsec and 
is rather soft (temperature of 0.3 keV). It extends to nuclear distances of 
more than 13 kpc and has a X-shaped appearance. Using simplifying assumptions
we derived parameters for this halo gas.

The X-ray luminosity of NGC~3079 is  higher by a factor of 10 compared to 
other galaxies of similar optical luminosity. It also exceeds the luminosity 
seen in starburst galaxies like NGC~253. We argue that this may be caused 
by the presence of an AGN in NGC~3079 rather then by starburst activity.
Investigations of the other galaxies in the NGC~3079 group, NGC~3073 and
MCG 9-17-9, showed that the X-ray emission is not extraordinary for
their type as might have been expected from the disturbance by the
galactic super-wind emanating from the active companion NGC~3079.  
 
\begin{acknowledgements}
This research has made use of the SIMBAD database operated at CDS, Strasbourg,
France and of the NASA/IPAC Extragalactic Database (NED) which is operated by
the Jet Propulsion Laboratory, CALTECH, under contract with the National
Aeronautics and Space Administration. To overlay the X-ray data we used an image
based on photographic data of the National Geographic Society -- Palomar
Observatory Sky Survey (NGS-POSS) obtained using the Oschin Telescope on
Palomar Mountain.  The NGS-POSS was funded by a grant from the National
Geographic Society to the California Institute of Technology.  The
plates were processed into the present compressed digital form with
their permission.  The Digitized Sky Survey was produced at the Space
Telescope Science Institute under US Government grant NAG W-2166.
GT acknowledges a Max-Planck Fellowship and the kind hospitality of
the ROSAT group during the completion of this project.  
The ROSAT project is supported by the German Bundesministerium f\"ur
Bildung, Wissenschaft, Forschung
und Technologie (BMBF/DLR) and the Max-Planck-Gesellschaft (MPG).
\end{acknowledgements}
 
\appendix
\section{Sources in the field of NGC~3079}
\label{appen}
\subsection{Correlation of HRI and PSPC detections}
\label{ap3}
\begin{table*}
\caption{X-ray properties of the sources in the NGC~3079 field and 
proposed identifications 
}
\label{coin}
{
\begin{flushleft}
\begin{tabular}{rrrrrlrrl}
\hline
\noalign{\smallskip}
ROSAT name &
Source &
$\Delta$ &
HRI flux &
PSPC flux &
Identification &$\Delta$
&log(${\rm f}_{\rm x} \over {\rm f}_{\rm v}$) &Ref.\\
\noalign{\smallskip}
(RX J)&no.
 &($''$)
&$^\ast$ &$^\ast$&& (\arcsec )&             & Notes \\
\noalign{\smallskip}
(1) &
(2) &
(3) &
(4) &
(5) &
(6) &
(7) &
(8) &
(9) \\
\noalign{\smallskip}
\hline
\noalign{\smallskip}
095956.3+554639 & P1 & & & 5.2$\pm$0.9 &APM O681 stellar (14.9, 0.5) & 17
& -2.2&1,6 \\
100008.8+553448 & P2 & & & 4.2$\pm$0.9&&&&6 \\
100032.3+553631& H1/P3 &
$2.1$ &
33.5$\pm$3.0 &
43.2$\pm$1.9  &APM O681 (18.0, 0.6) & 4 & -0.1 &1 \\
100037.4+555046& P4 & &$<$12 & 3.1$\pm$0.8 &APM 1331 stellar (20.0, 0.3) & 6 &
-0.2 &1 \\
100056.4+553519&H2/P5 &
$3.0$ &
17.6$\pm$2.0 &
19.4$\pm$1.3 &
APM O681 stellar (19.6, 0.6) & 2 & 0.3&1 \\
&&&& & radio source & 5 &
&2 \\
100058.9+555141 &H3 & &3.8$\pm$1.1 &$<$2.3  &
APM 1331 (20.5, 2.4) & 6 & -0.8
&1 \\
100104.7+553519 &H4/P6 &
$3.0$ &
3.6$\pm$1.1 &
5.3$\pm$0.8 &
APM O681 stellar (18.9, 0.7) & 6 & -0.6
&1 \\
& &&&&Q0957+5549  z=1.5 & 4 &
&3,5 \\
100110.0+552838&H5/P7 &
$11.4$ &
($>$9.4$\pm$2.0) &
19.7$\pm$1.4 &
APM 1331 stellar (16.9, 0.3) & 5 & -0.7
&1,6 \\
&&&& &Q0957+5543 z=2.1 & 4 &
&3,5 \\
100113.3+554906& P8 & & $<$0.7 & 3.7$\pm$0.9 \\
100114.6+554310&H6/P9 &
$0.7$ &
3.3$\pm$0.9 &
4.4$\pm$0.7 &
APM 1331 (8.9, 1.3) & 7 & -5.0
&1  \\
&&&& &MCG 9-17-9 & 7 &
&2 \\
100116.8+554709 &H7     & & 2.1$\pm$0.8 & $<$2.1 \\
100119.7+554559 &H8/P10 &
$5.3$ &
5.0$\pm$1.1 &
10.0$\pm$1.0&
APM 1331 stellar (21.7, $<$1.7) & 1 & $>$0.4
&1 \\
100120.9+553244 & P11 & & $<$2.8 & 1.9$\pm$0.6 \\
100120.6+555355 &H9/P12 &
$6.3$ &
146.1$\pm$5.6 &
200.8$\pm$3.8 &
APM 1331 (14.3, 0.1), & 4 & -0.6
&1 \\
 &&&& &
lensed Q0957+5608AB z=1.4 & 3 &
&2,5 \\
100123.5+553953 &H10 &      & 1.7$\pm$0.7 & $<$1.6 &\\
100130.2+554033 & P13 & & $<$0.5 & 3.6$\pm$0.9 & \\
100138.2+553508 &H11 & & 2.3$\pm$0.8 & $<$2.3 &
HD 86661 (8.7, 0.7) G8IV-V$^{\ast \ast}$ & 1 & -5.1
&4 \\
100146.6+555036 & P14 & & $<$1.6 & 2.2$\pm$0.6 \\
100156.3+555434& P15 & & $<$2.2 & 4.2$\pm$0.8 &
APM 1331 stellar (18.7, 0.9)  & 4 & -0.9
&1 \\
100200.5+553200  & P17 & & $<$1.9 & 7.4$\pm$0.9 &
APM 1331 stellar (20.8, 0.9)  & 10 & 0.2
&1  \\
100205.0+554555 &H17      & & 1.5$\pm$0.6 & $<$5.3 \\
100206.2+552750 &P20 & & $<$1.6 & 1.2$\pm$0.6 \\
100223.0+553429 &H19/P22 &
$9.0$ &
6.7$\pm$1.3 &
5.7$\pm$0.8 &
HD 237859 (10.6, 1.1) F0 & 2 & -4.0
&4  \\
100228.2+553941 & P23 & & $<$0.3 & 2.4$\pm$0.7 &  \\
100237.4+553505 & P24 & & $<$3.7 & 2.0$\pm$0.6 &
APM 1331 stellar (19.7, 3.7) & 13 & -1.9
&1 \\
100240.8+552750 & P25 & & $<$1.6 & 2.8$\pm$0.7&
APM 1331 (21.7, 3.1) & 14 & -0.7
&1 \\
100245.5+554648 &H20/P26 &
$5.8$ &
6.2$\pm$1.2 &
4.3$\pm$0.7&
APM 1331 stellar (21.8, $<$1.8)  & 4 & $>$0.0
&1  \\
100244.9+555110 & P27 & & $<$1.2 & 2.2$\pm$0.7&
radio source & 13 &
&2  \\
100245.1+555758 & P28 & &  & 17.3$\pm$1.5&&&&6 \\
100254.2+554226 &H21   & & 2.3$\pm$0.8 & $<$1.3 \\
100303.4+554752 &H22/P29 &
$5.5$ &
3.9$\pm$1.1 &
3.0$\pm$0.7&
APM 1331 stellar (12.2, 0.9) & 3 & -3.6
&1,6  \\
100305.5+553727 & P30 & & $<$1.5 & 1.6$\pm$0.6 \\
100309.6+554135 &H23/P31 &
$5.2$ &
14.6$\pm$1.9 &
14.6$\pm$1.2 &
APM 1331 stellar (19.1, 0.5) & 1 & 0.0
&1 \\
100327.8+552625 & P32 & &$^{\ast\ast\ast}$ & 3.7$\pm$0.9 &
APM 1331 stellar (20.4, 0.9) & 9 & -0.2
&1  \\
100333.5+553627 & P33 & &$<$8  & 5.4$\pm$0.9 \\
100343.6+553331 & P34 & &  & 4.9$\pm$0.9&&&&6 \\
\noalign{\smallskip}
\hline
\end{tabular}
\end{flushleft}
\[
\begin{array}{lp{0.95\linewidth}}
$$^\ast$$ & fluxes or $2\sigma$ upper limits in units of 
10$^{-14}$\,\ergcm\
for a 5~keV thermal bremsstrahlung spectrum in the 0.1--2.4~keV
band, corrected for Galactic absorption\\
$$^{\ast \ast}$$ & The optical position has been corrected for
proper motion\\
$$^{\ast \ast \ast}$$ & outside HRI field of view
\end{array}
\]
References and Notes: (1) Irwin \et 1994; (2) Irwin \& Seaquist 1991; (3) Bowen
\et 1994; (4) SIMBAD; (5) NED; (6) source at an off-axis angle $\ge 15'$
In the HRI detector.  No upper limit is calculated for these sources
(see text) \\
}
\end{table*}

Thirty-four sources are detected around NGC~3079 in the PSPC and HRI
observations.  Table~\ref{coin} shows their characteristics.  If the
source has only been detected in one of the detectors, a $2\sigma$
upper limit at the source position has been determined for the other
detector.  No upper limit is computed for sources detected in the PSPC
that are at a large off-axis angle in the HRI, since we cannot reliably
measure the count rates (the effective exposure time could be lowered as a
consequence of the wobbling motion that can place the source outside
the field of view for part of the observation, the detector sensitivity
and the PSF are poorly known). 
We checked the mean offset of the source positions (listed
in col. 3 of Table~\ref{coin} for those sources detected in both
instruments) to verify our attitude solution of Sect.~\ref{sdet-anal}
for all sources.  The mean offsets in the north-south and east-west
directions were below the systematical error.

The distribution of the hardness ratios listed in Table~\ref{sources_pspc} 
is shown in Fig.~\ref{hr-theor}.  Comparison with the theoretical values
calculated for power law and thermal brems\-strah\-lung spectra
(Fig.~\ref{hr-theor})  show that in most cases the sources have hardness
ratios consistent with a moderate amount of absorption, as expected from 
the line-of-sight Galactic \HI\ column density value.  

For several  sources the derived fluxes or upper limits were comparable in the
two separate observations.  For four sources detected in both
observations, namely H1/P3, H5/P7, H8/P10, and H9/P12 the flux
differences exceed 2$\sigma$.  For one additional source
detected only in one of the observations, namely  P17,
the 2$\sigma$ HRI upper limit is lower by more than twice the error on the
corresponding detection.  
Since the expected uncertainty in the count
to flux conversion factor is of the order of 15\% at most (see
Sect.~\ref{observations}), the assumption of a wrong spectral model
should not cause these flux differences.  However, it should be noted
that two of these sources, H1/P3 and H5/P7, are at
large off-axis angles,
so possible effects due to their positioning in the detector might come
into play.   We have checked for example the position of source H5 in
detector coordinates, to understand whether the wobbling motion could
affect the effective exposure of this source, and found that it is in
fact possible that the source falls outside of the detector field of view
for part of the observation, so the flux quoted should in fact be
considered a lower limit.  The same consideration does not apply to the
other sources and it is therefore likely that these sources have
varied between the two observations.  As discussed in the next section,
four of the variable sources are coincident in position with stellar
objects or active nuclei (see Table~\ref{coin}), which as a class 
are known to be
variable sources, and would therefore support the idea that the
observed variations are real: H5/P7 is identified with QSO 0957+5543 at
a redshift z=2.1, H9/P12 with the well studied QSO 0957+5608AB, a
double source originating from a quasar at z=1.4 that is
gravitationally lensed by a cluster of galaxies at z=0.4.

\subsection{Identification of X-ray sources in the field of NGC~3079}
For each source detected we prepared
APM finding charts (Irwin et al. 1994), from which we can get a positional
accuracy for the optical candidates of better than 2\arcsec,
optical magnitudes in blue (O) and red (E),  and the color index O-E. In the
charts sources are classified as stellar, non-stellar, or blend. 
Optical counterparts were searched for each in a circle of the X-ray 
position error.

Several positional coincidences were found from these charts.  
Additional candidates resulted from correlations of our source catalog with 
the entries in the
SIMBAD and NED databases and with radio sources in the field given in 
Irwin and Seaquist (1991).  

To improve on the reliability of the identifications, 
we have estimated the log(f$_{\rm x}$/f$_{\rm v}$) = log(f$_{\rm x}$) + 
(m$_{\rm v}$/2.5) + 5.37 (see Maccacaro et al. 1988), that we can use
to discriminate between stars, for which log(f$_{\rm x}$/f$_{\rm v}$)
is generally $<$ -1, and  AGN for which this quantity is in the range  -1.2 
to +1.2 (see also Brinkmann 1992 and Pfl\"uger et al. 1996). 
X-ray fluxes in the 0.1--2.4 keV band have been
determined from the count rates assuming a 5 keV thermal brems\-strah\-lung 
spectrum, corrected for Galactic absorption (see Table~\ref{conversion}).  
For the
optical magnitude we used the 'E' magnitude from the APM catalog.
  
Proposed identifications for 21 of the ROSAT field sources are summarized 
in Table~\ref{coin}.   Column 6  lists the
APM field and  classification, and in parenthesis the O luminosity and
the color (O--E).   The optical identification and the 
spectral type given for stars are taken from the SIMBAD 
database.   Identifications for extragalactic objects, with their
redshift when available, is based on the NED database.  
The separation to the X-ray position is given in col. 7. 
log(f$_{\rm x}$/f$_{\rm v}$) and references used 
are indicated in cols. 8 and 9.

In most cases there is only one APM candidate within the error radius of
the X-ray source. 
 All proposed identifications are located within the X-ray position errors. 
Three sources are identified with known QSOs (H4/P6, H5/P7, H9/P12). 
Two sources (H2/P5 and P27) are identified
with radio sources, for the first of the two there is also a APM candidate.
Two sources (H11 and (H19/P22)) are identified with stars cataloged in SIMBAD.

The log(f$_{\rm x}$/f$_{\rm v}$) values and the
optical colors of the unidentified objects show values similar to those
of the identified QSOs.  This suggest that most of them are also
AGN/QSOs. 

log(f$_{\rm x}$/f$_{\rm v}$) $<$-1 are seen in only three unidentified
sources, P1, P24  and  H22/P29, for which identifications with stars
are suggested.  
The optical candidates of H3, P25, and P24 show rather red colors. 
The candidates for H3 and P25 are flagged by APM
as non-stellar in the red and stellar
in the blue.   It is likely that an AGN embedded in a host galaxy is
the optical counterpart of these sources.

\end{document}